\documentclass[prb,twocolumn,nofootinbib,citeautoscript,10pt,notitlepage]{revtex4-2}
\usepackage{dsfont}
\usepackage{float}
\usepackage{array}
\usepackage{slashed,bbold}

\usepackage{amsmath,amssymb,bm} 
\usepackage{graphicx}

\usepackage[tight]{subfigure} 

\usepackage{color} 
\usepackage[papersize={8.5in,11in}]{geometry}

\usepackage{color}
\definecolor{darkblue}{rgb}{0.,0.,0.4}
\definecolor{darkred}{rgb}{0.5,0.,0.}
\definecolor{BlueViolet}{RGB}{138,43,226}
\definecolor{SkyBlue}{RGB}{30,144,255}
\definecolor{DarkGreen}{RGB}{0,100,0}
\usepackage[pdftex,colorlinks=true,linkcolor=darkblue,citecolor=blue,urlcolor=darkred]{hyperref}

\def \be{\begin{align}}
\def \ee{\end{align}}
\def \bea{\begin{eqnarray}}
\def \eea{\end{eqnarray}}
\def \nn{\nonumber \\}

\begin{document}

\title{Valley-polarized nematic order in twisted moir\'e systems: In-plane orbital magnetism and non-Fermi liquid to Fermi liquid crossover}

\author{Ipsita Mandal}

\affiliation{Institute of Nuclear Physics, Polish Academy of Sciences, 31-342 Krak\'{o}w,
Poland~\\
Department of Physics, Stockholm University, AlbaNova University
Center, 106 91 Stockholm, Sweden}

\author{Rafael M. Fernandes}

\affiliation{ School of Physics and Astronomy, University of Minnesota, Minneapolis,
Minnesota 55455, USA }

\begin{abstract}
The interplay between strong correlations and non-trivial topology
in twisted moir\'e systems can give rise to a rich landscape of ordered
states that intertwine the spin, valley, and charge degrees of freedom.
In this paper, we investigate the properties of a system that displays long-range valley-polarized nematic order. Besides breaking the threefold rotational symmetry of the triangular moir\'e
superlattice, this type of order also breaks twofold rotational and
time-reversal symmetries, which leads to interesting properties. First, we develop a phenomenological model to describe the onset of this ordered state in twisted moir\'e systems, and to explore its signatures in their thermodynamic and electronic properties. Its main manifestation is that it triggers the emergence of in-plane orbital magnetic moments oriented along high-symmetry lattice directions. We also investigate the properties of the valley-polarized nematic state at zero temperature. Due to the existence of a dangerously irrelevant coupling $\lambda$ in the six-state clock model that describes the putative valley-polarized nematic quantum critical point, the ordered state displays a pseudo-Goldstone mode. Using a two-patch model, we compute the fermionic self-energy to show that down to very low energies, the Yukawa-like coupling between the pseudo-Goldstone mode and the electronic degrees of freedom promotes the emergence of non-Fermi liquid behaviour. Below a crossover energy scale $\Omega^{*}\sim\lambda^{3/2}$, however, Fermi liquid behaviour is recovered. Finally, we discuss the applicability of these results to other non-trivial nematic states, such as the spin-polarized nematic phase.
\end{abstract}
\maketitle

\tableofcontents

\section{Introduction}

The observation of electronic nematicity in the phase diagrams of twisted bilayer graphene \cite{Jiang2019,Kerelsky2019,Choi2019,Cao2021} and twisted double-bilayer graphene \cite{RubioVerdu2021,Samajdar2021} provides a new setting to elucidate these electronic liquid crystalline states, which spontaneously break the rotational symmetry of the system. Shortly after nematicity was proposed to explain certain properties of high-temperature
superconductors \cite{Kivelson98}, it was recognized that the Goldstone mode of an ideal electronic nematic phase would have a profound impact on the electronic properties of a metal \cite{Oganesyan01,Kim2004,Garst09}. This is because, in contrast to other Goldstone modes such as phonons and magnons, which couple to the electronic density via a gradient term, the nematic Goldstone mode displays a direct Yukawa-like coupling to the electronic density \cite{Ashvin14}. As a result, it is expected to promote non-Fermi liquid (NFL) behaviour, as manifested
in the sub-linear frequency dependence of the imaginary part of the electronic self-energy \cite{Oganesyan01,Garst2010}.

However, because the crystal lattice breaks the continuous rotational symmetry of the system, the electronic nematic order parameter realized in layered quantum materials has a discrete $Z_{q}$ symmetry, rather than the continuous XY (or O(2)) symmetry of its two-dimensional (2D) liquid crystal counterpart \cite{Fradkin_review}. In the square lattice, the $Z_{2}$ (Ising-like) symmetry is associated with selecting one of the two orthogonal in-plane directions, connecting either nearest-neighbor or next-nearest-neighbor
sites \cite{Fernandes2014}. In the triangular lattice, the $Z_{3}$ (three-state Potts/clock) symmetry refers to choosing one of the three bonds connecting nearest-neighbor sites \cite{Hecker2018,Fernandes_Venderbos}. In both cases, the excitation spectrum in the ordered state
is gapped, i.e. there is no nematic Goldstone mode. Consequently, NFL behaviour is not expected to arise inside the nematic phase --- although it can still emerge in the disordered state due to interactions mediated by possible quantum critical fluctuations \cite{Metzner03,Rech2006,max-subir,Lee-Dalid,ips-lee,ips-uv-ir,ips-subir,max-cooper,ips-sc,Lederer2015,Klein2018}.

In twisted moir\'e systems \cite{Andrei2020_review,Balents2020_review}, which usually display an emergent triangular moir\'e superlattice, another type of nematic order can arise due to the presence of the valley degrees of freedom: a \emph{valley-polarized nematic state} \cite{cenke}. Compared with the standard nematic state, valley-polarized nematic order breaks not only the threefold rotational symmetry of the lattice, but also ``inversion'' (more precisely, two-fold rotational), and time-reversal symmetries. It is another example, particularly relevant for moir\'e superlattices, of a broader class of ``non-standard" electronic nematic orders that are intertwined with additional symmetries of the system, such as the so-called nematic spin-nematic phases \cite{Kivelson_RMP,Wu_Fradkin2007,Fischer_Kim2011}. 

In twisted bilayer graphene (TBG) \cite{cao-insulator,cao,Yankowitz2019,Efetov19}, while threefold rotational symmetry-breaking \cite{Jiang2019,Kerelsky2019,Choi2019,Cao2021} and time-reversal symmetry-breaking \cite{Sharpe19,Efetov19,Young19,Tschirhart2021} have been observed in different regions of the phase diagram, it is not clear yet whether a valley-polarized nematic state is realized. Theoretically, the valley-polarized nematic order parameter has a $Z_{6}$ symmetry, which corresponds to the six-state clock model \cite{Jose1977}. Interestingly, it is known that the six-state clock model transition belongs to the XY universality class in three spatial dimensions, with the sixfold anisotropy perturbation being irrelevant at the critical point \cite{Amit1982,Oshikawa2000,Sudbo2003,fucito}.  

Thus, at $T=0$ and in a 2D triangular lattice, a valley-polarized nematic quantum critical point (QCP) should share the same universality class as a QCP associated with a hypothetical XY electronic nematic order parameter, that is completely decoupled from the lattice degrees of freedom \cite{Shibauchi2020}. In other words, a 2D 6-state clock model exhibits a continuous phase transition at $T=0$, that is described by a $(2+1)\rm{D}$ Ginzburg-Landau theory of an O(2) order parameter, with a $Z_6$ anisotropic term --- the latter is irrelevant in the renormalization group (RG) sense. In fact, the sixfold anisotropy term is \emph{dangerously irrelevant} \cite{Amit1982}, as it becomes a relevant perturbation inside the ordered state \cite{Oshikawa2000,Sandvik2007,Okubo2015,Leonard2015,Podolski2016,Sandvik2020}. As a result, the valley-polarized nematic phase displays a pseudo-Goldstone mode, i.e., a would-be Goldstone mode with a small gap, that satisfies certain scaling properties as the QCP is approached \cite{Sandvik2021}. 

In this paper, we explore the properties of the valley-polarized nematic state in twisted moir\'e systems, and, more broadly, in a generic metal. We start from a phenomenological SU$(4)$ model, relevant for TBG, which is unstable towards intra-valley nematicity. We show that, depending on the inter-valley coupling, the resulting nematic order can be a ``standard" nematic phase, which only breaks threefold rotational symmetry, or the valley-polarized nematic phase, which also breaks twofold and time-reversal symmetries. By employing group-theory techniques, we show that the onset of valley-polarized nematicity triggers in-plane orbital magnetism, as well as standard nematicity and different types of order in the valley degrees of freedom. The $Z_6$ symmetry of the valley-polarized nematic order parameter is translated as six different orientations for the in-plane magnetic moments. Moving beyond phenomenology, we use the six-band tight-binding model for TBG of Ref. \cite{Po2019} to investigate how valley-polarized nematic order impacts the electronic spectrum. Because the combined $C_{2z} \mathcal{T}$ symmetry is preserved, the Dirac cones remain intact, albeit displaced from the $K$ point. Moreover, band degeneracies associated with the valley degrees of freedom are lifted, and the Fermi surface acquires characteristic distortion patterns.

We next study the electronic properties of the valley-polarized nematic phase at $T=0$, when a putative quantum critical point is crossed. To make our results more widely applicable, we consider the case of a generic metal with a simple circular Fermi surface. First, we show that the phase fluctuations inside the valley-polarized phase couple directly to the electronic density. Then, using a two-patch model \cite{max-subir,Lee-Dalid,ips-lee,ips-uv-ir,ips-fflo,ips-nfl-u1}, we show that the electronic self-energy $\Sigma$ displays, along the hot regions of the Fermi surface and above a characteristic energy $\Omega^{*}$, the same NFL behaviour as in the case of an ``ideal'' XY nematic
order parameter \cite{Oganesyan01,Garst2010}, i.e. $\Sigma\left(\nu_{n}\right)\sim i\left|\nu_{n}\right|^{2/3}$, where $\nu_{n}$ is the fermionic Matsubara frequency. Below $\Omega^{*}$, however, we find that $\Sigma\left(\nu_{n}\right)\sim i\,\nu_{n}$, and Fermi liquid (FL) behaviour is restored. Moreover, the bosonic self-energy,
describing the phase fluctuations, acquires an overdamped dynamics due to the coupling to the fermions. 

Exploiting the scaling properties of the six-state clock model, we argue that this NFL-to-FL crossover energy scale $\Omega^{*}$, which is directly related to the dangerously irrelevant variable $\lambda$ of the six-state clock model via $\Omega^{*}\sim\lambda^{3/2}$, is expected to be much smaller than the other energy scales of the problem. As a result, we expect NFL behaviour to be realized over an
extended range of energies. We discuss possible experimental manifestations of this effect at finite temperatures, and the extension of this mechanism to the case of \emph{spin-polarized nematic order} \cite{Wu_Fradkin2007}, which has been proposed to occur in moir\'e systems with higher-order Van Hove singularities \cite{Classen2020,Chichinadze2020}. We also discuss possible limitations of the results arising from the simplified form assumed for the Fermi surface.

The paper is organized as follows: Sec.~\ref{sec_phenomenology} presents a phenomenological description of valley-polarized nematic order in TBG, as well as its manifestations on the thermodynamic and electronic properties. Sec.~\ref{secmodel} introduces the bosonic and fermionic actions that describe the system inside the valley-polarized nematic state. Sec.~\ref{secnfl} describes the results for the electronic self-energy, obtained from both the Hertz-Millis theory and the patch methods, focussing on the onset of an NFL behaviour. In Sec.~\ref{secend}, we discuss the implications of our results for the observation of NFL behaviour in different types of systems. 


\section{Valley-polarized nematic order in TBG} \label{sec_phenomenology}

\subsection{Phenomenological model}

In TBG, the existence of electron-electron interactions larger than the narrow bandwidth of the moir\'e bands \cite{BM_model,Tarnopolsky2019} enables the emergence of a wide range of possible ordered states involving the spin, valley, and sublattice degrees of freedom \cite{Rademaker2018,Isobe2018,Kennes2018,Venderbos18,Sherkunov2018,Thomson2018,Kang2019,Seo2019,Yuan2019_vhs,Bascones19,Natori2019,Vafek2020,cenke,Bultinck2020,Xie2020,Cea2020,Christos2020,Fernandes_ZYMeng,Bernevig_TBGVI,Brillaux2020,ips-tbg,Chichinadze2021,Song2021}. Here, we start by considering a model for TBG that has U(1) valley symmetry.
Together with the symmetry under independent spin rotations on the
two valleys, the model has an emergent SU(4) symmetry, and has been
widely studied previously \cite{Kang2019,Bultinck2020,Vafek_Kang_PRL_2020,Bernevig_TBGVI,Wang_Kang_RMF,Chichinadze2021}. Within the valley subspace $a$, we assume
that the system has an instability towards a nematic phase, i.e. an
intra-valley Pomeranchuk instability that breaks the $C_{3z}$ rotational
symmetry. Indeed, several models for TBG have found proximity
to a nematic instability \cite{Dodaro2018,Vafek2020,cenke,Bernevig_TBGVI,Brillaux2020,Nori2020,Chichinadze2020,Khalaf2020,Kontani2022}. Note here that $a=+,-$ refers to the moir\'e
valley. Hereinafter, we assume that the valleys are exchanged by a $C_{2x}$
rotation. Let the intra-valley nematic order associated with valley
$a$ be described by the two-component order parameter $\boldsymbol{\varphi}_{a}=\left(\varphi_{a,1},\,\varphi_{a,2}\right)$
that transforms as the $\left(d_{x^{2}-y^{2}},d_{xy}\right)$-wave
form factors. 

A single valley does not have $C_{2z}$ symmetry or $C_{2x}$ symmetry
(but it does have $C_{2y}$ symmetry); it is the full system, with
two valleys, that has the symmetries of the $D_{6}$ space group.
A $C_{2z}$ rotation (or, equivalently, a $C_{2x}$ rotation) exchanges
valleys $+$ and $-$. Time-reversal $\mathcal{T}$ has the same effect.
If the valleys were completely decoupled, the nematic free energy
would be
\begin{align}
F_{0}\left(\boldsymbol{\varphi}_{+},\boldsymbol{\varphi}_{-}\right)=
r_{0}\left(\boldsymbol{\varphi}_{+}^{2}
+ \boldsymbol{\varphi}_{-}^{2}\right)+\mathcal{O}\left(\boldsymbol{\varphi}_{a}^{3}\right),
\label{F_0}
\end{align}
to leading order.
However, since independent spatial rotations on the two valleys are
not a symmetry of the system, there must be a quadratic term coupling
the two intra-valley nematic order parameters, of the form
\begin{align}
\bar{F}=\kappa\left(\boldsymbol{\varphi}_{+}\cdot\boldsymbol{\varphi}_{-}\right)
=\frac{\kappa}{2} \, \boldsymbol{\varphi}_{a}\cdot\tau_{aa'}^{x}
\,\boldsymbol{\varphi}_{a'}\,,
\label{delta_F}
\end{align}
where $\tau^{i}$ is a Pauli matrix in valley space. This term is
invariant under both $C_{2z}$ and $\mathcal{T}$, as it remains the
same upon exchange of the two valleys. Moreover, it is invariant under
$C_{3z}$, since it is quadratic in the nematic order parameters. It
is important to note that $C_{3z}$ must be considered as a global
threefold rotation, equal in both valleys.

Minimizing the full quadratic free energy, we find two possible orders
depending on the sign of $\kappa$, which ultimately
can only be determined from microscopic considerations. For $\kappa<0$,
the resulting order parameter
\begin{align}
\tilde{\boldsymbol{\Phi}}=\boldsymbol{\varphi}_{+}+\boldsymbol{\varphi}_{-}
\end{align}
is valley-independent.
It has the same transformation properties as $\boldsymbol{\varphi}_{a}$
under $C_{3z}$, and it is even under both $C_{2z}$ and $\mathcal{T}$.
As a result, it must transform as the $E_{2}^{+}$ irreducible representation of $D_{6}$
(the ``$+$'' superscript indicates that it is even under time-reversal).
This is the usual nematic order parameter, which belongs to the three-state
Potts/clock model universality class. Indeed, parametrizing $\tilde{\boldsymbol{\Phi}}=\tilde{\Phi}_0\left(\cos\tilde{\alpha},\,\sin\tilde{\alpha}\right)$,
one finds a free-energy
\begin{align}
\tilde{F}=r \, \tilde{\Phi}_0^{2}
-2 \, \lambda \, \tilde{\Phi}_0^{3} \cos(3\tilde{\alpha}) + u \,\tilde{\Phi}_0^{4}\,,
\end{align}
corresponding to the three-state Potts or clock
model \cite{Fernandes_Venderbos,cenke}.
For $\kappa>0$, the resulting order parameter
\begin{align}
\boldsymbol{\Phi}=\boldsymbol{\varphi}_{+}-\boldsymbol{\varphi}_{-}
\end{align}
is valley-polarized.
The key difference between this phase and the earlier one is
that $\boldsymbol{\Phi}$ is odd under both $C_{2z}$ and $\mathcal{T}$,
while retaining the same transformation properties under $C_{3z}$.
Therefore, $\boldsymbol{\Phi}$ must transform as the $E_{1}^{-}$
irreducible representation of $D_{6}$, with the ``$-$'' superscript indicating that it is
odd under time-reversal. This is the valley-polarized nematic order
parameter, first identified in Ref. \cite{cenke}. The full free-energy for $\boldsymbol{\Phi}$ can be obtained
from its symmetry properties, rather than starting from the uncoupled
free energies in Eq.~\eqref{F_0}. Parametrizing $\boldsymbol{\Phi}=\Phi_0 \left(\cos\alpha,\,\sin\alpha\right)$,
one finds the following free-energy expansion \cite{cenke}:

\begin{align}
F=r \, \Phi_0^{2}+u \,\Phi_0^{4}-2 \, \lambda\, \Phi_0^{6} \cos(6 \alpha) \,.
\label{eq:F6}
\end{align}

The $\lambda$-term in Eq.~\eqref{eq:F6} is the lowest-order term that lowers the symmetry
of $\boldsymbol{\Phi}$ from O(2) to $Z_{6}$. As a result, the action
corresponds to a six-state clock model. Indeed, minimization of the
action with respect to the phase $\alpha$ leads to six different
minima, corresponding to (1) $\alpha=\frac{\pi}{3}\,n$ for $\lambda>0$;
and (2) $\alpha=\frac{\pi}{3}\left(n+\frac{1}{2}\right)$ for $\lambda<0$
(with $n=0,\ldots,5$). At finite temperatures, the 2D six-state
clock model undergoes two Kosterlitz-Thouless transitions: the first
one signals quasi-long-range order of the phase $\alpha$, whereas
the second one marks the onset of discrete symmetry-breaking and long-range
order \cite{Jose1977}.

\subsection{Manifestations of the valley-polarized phase}

The onset of valley-polarized order leads to several observable consequences.
First, we note that the in-plane magnetic moment $\mathbf{m}=\left(m_{x},\,m_{y}\right)$
also transforms as $E_{1}^{-}$. Therefore the following linear-in-$\Phi$
free-energy coupling term is allowed:
\begin{align}
\delta F_{1}\sim\mathbf{m}\cdot\boldsymbol{\Phi}  \,.  
\end{align}

This implies that valley-polarized nematic order necessarily triggers
in-plane magnetic moments --- see also Ref. \cite{Berg2022} for the case of in-plane magnetic moments induced by hetero-strain. These moments are directed towards the angles $\alpha$ that minimize the sixth-order term $\Phi_0^{6}\cos (6\alpha)$
of the nematic free energy. Because the system has SU(2) spin-rotational
invariance, $\mathbf{m}$ must be manifested as an in-plane orbital
angular magnetic moment. Therefore, valley-polarized nematic order provides
a mechanism for in-plane orbital magnetism, which is complementary
to previous models for out-of-plane orbital magnetism.

There are additional manifestations coming from higher-order terms
of the free energy. Valley-polarized nematic order $\boldsymbol{\Phi}$
also induces the ``usual'' nematic order $\tilde{\boldsymbol{\Phi}}$
via the quadratic-linear coupling
\begin{align}
\delta F_{2}\sim \left(\Phi_{1}^{2}-\Phi_{2}^{2}\right)
\tilde{\Phi}_{1}-2\Phi_{1} \, \Phi_{2}\, \tilde{\Phi}_{2}
=\Phi_0^{2}\, \tilde{\Phi}_0
\cos\left(2\alpha+\tilde{\alpha}\right).
\end{align}
Moreover, $\boldsymbol{\Phi}$ also induces either the order parameter
$\eta$, which transforms as $B_{2}^{-}$, or the order parameter
$\tilde{\eta}$, which transforms as $B_{1}^{-}$. Both $\eta$ and
$\tilde{\eta}$ are even under $C_{3z}$, but odd under $C_{2z}$ and
$\mathcal{T}$. The only difference is that $\eta$ is odd under $C_{2x}$
and even under $C_{2y}$, whereas $\tilde{\eta}$ is odd under $C_{2y}$
and even under $C_{2x}$. We find the following cubic-linear terms are allowed:
\begin{align}
\delta F_{3}^{(1)} & \sim\left(3\,\Phi_{1}^{2} \, \Phi_{2}-\Phi_{2}^{3}\right)\eta
=\Phi_0^{3} \,\eta\,\sin (3\alpha) \,,\nn
\delta F_{3}^{(2)} & \sim
\left(\Phi_{1}^{3}-3 \, \Phi_{1} \,\Phi_{2}^{2}\right)\tilde{\eta}
=\Phi_0^{3}\,\tilde{\eta}\,\cos(3\alpha)\,.
\end{align}
Since
\begin{align}
\cos^{2}(3\alpha)  =\frac{1+\cos (6\alpha)}{2} \text{ and }
\sin^{2}(3\alpha)  =\frac{1-\cos (6\alpha)}{2} \,,
\end{align}
we conclude that, if the coefficient $\lambda$ of the $\Phi_0^{6}\cos (6\alpha)$
term is positive [implying $\cos (6\alpha)=+1$], $\tilde{\eta}$  is
induced. Otherwise, if $\lambda$ is negative [implying $\cos (6\alpha)=-1$], $\eta$ is induced.
Physically, $\eta$ can be interpreted as a valley charge polarization
$\eta=\rho_{+}-\rho_{-}$, where $\rho_{a}$ is the charge at valley
$a$. That is because $C_{2x}$ also switches valleys $1$ and $2$.
On the other hand, $C_{2y}$ does not involve valley switching and
is therefore an intra-valley type of order.

\subsection{Impact of the valley-polarized order on the electronic spectrum}

 \begin{figure}[]
   \centering
\subfigure[]{\includegraphics[width= 0.4\textwidth]{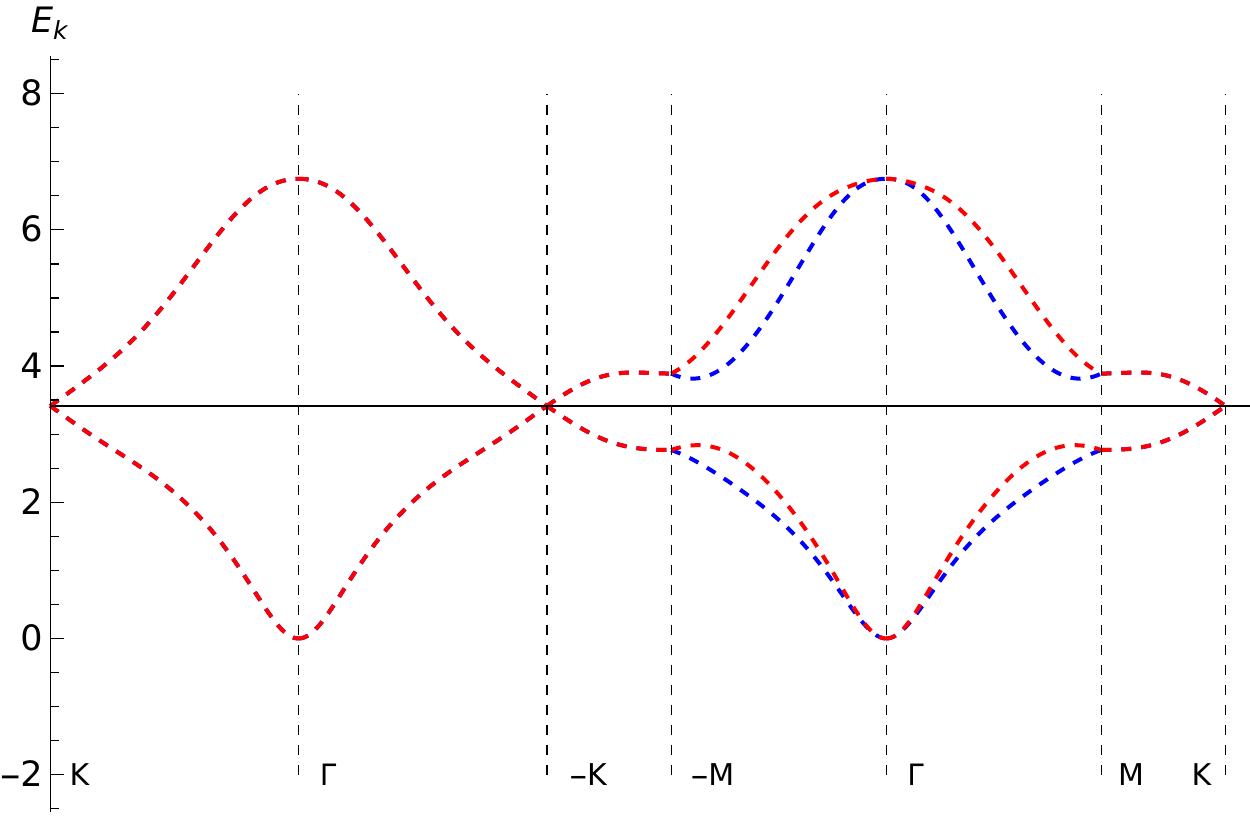}}
\subfigure[]{\includegraphics[width= 0.4 \textwidth]{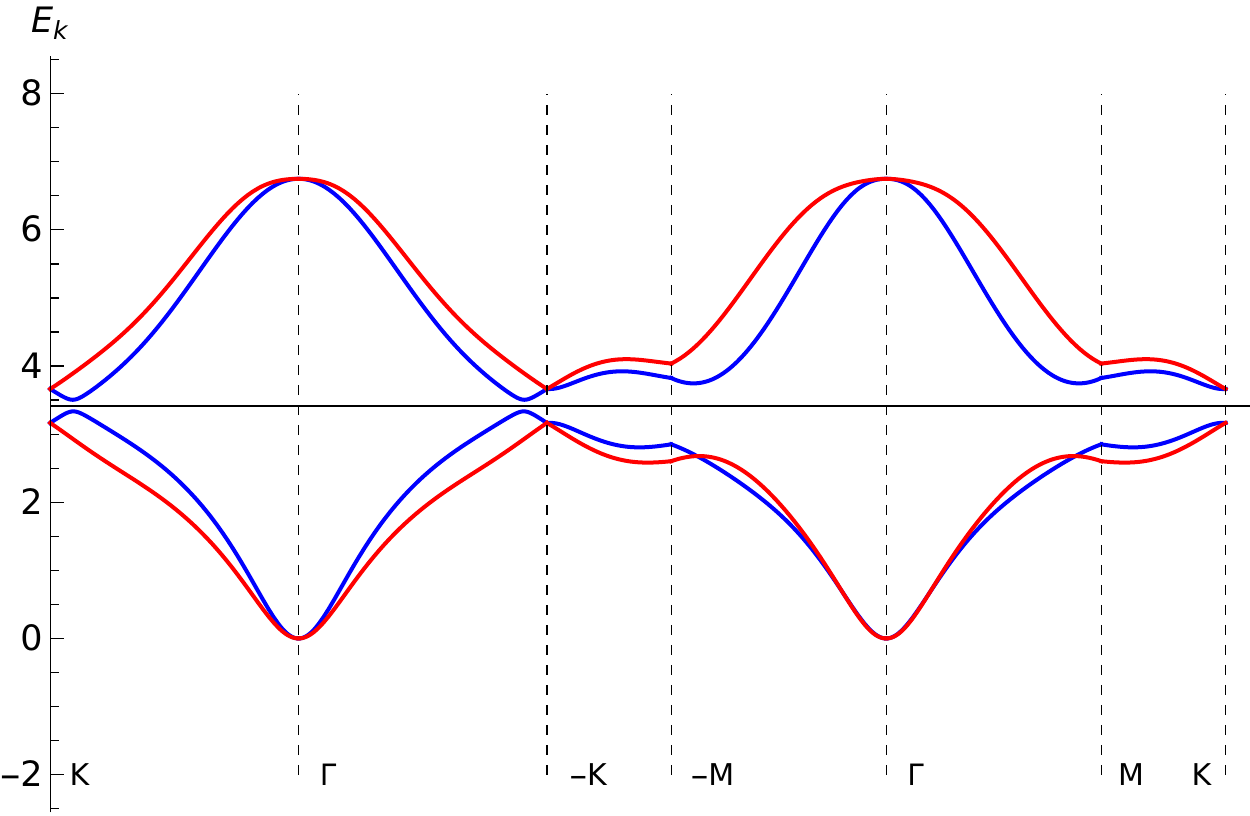}}
\caption{\label{fig:bands}
Band structure along the high symmetry directions of the moir\'e Brillouin zone, for the (almost) flat bands of TBG. This is numerically computed from the six-orbital model of Ref.~\cite{Po2019}, without [panel (a); dashed lines] and with [panel (b); solid lines] valley-polarized nematic ordering. Red and blue lines refer to the two valleys. The parameters used are the same as in Ref.~\cite{Po2019}, and we have chosen $\Phi_0=0.01 \,t_{\kappa}$ and $\alpha=0$ for the ordered state. The energy values shown are in meV.}
 \end{figure}

 \begin{figure*}
   \centering
\subfigure[]{\includegraphics[width= 0.15\textwidth]{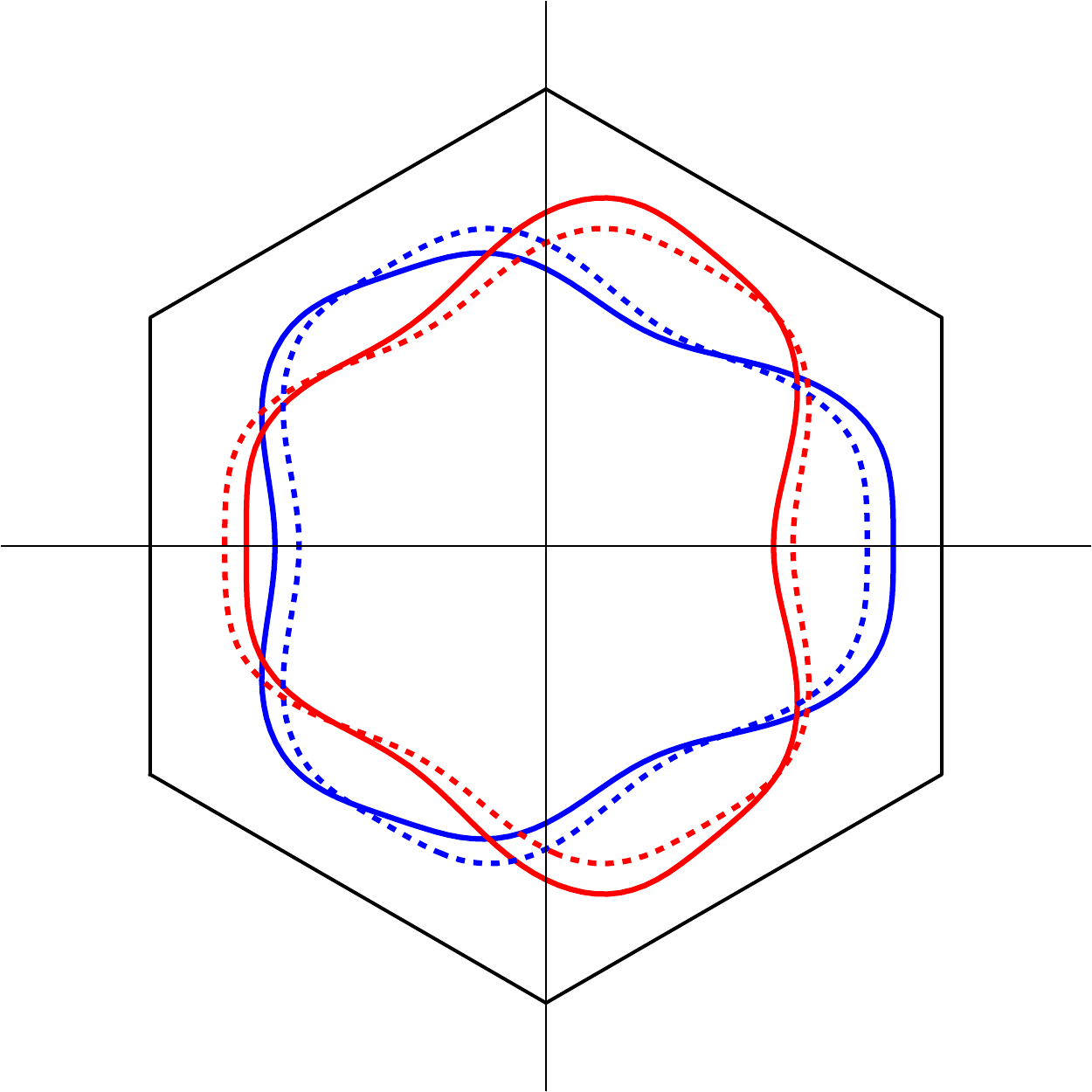}}
\subfigure[]{\includegraphics[width= 0.15\textwidth]{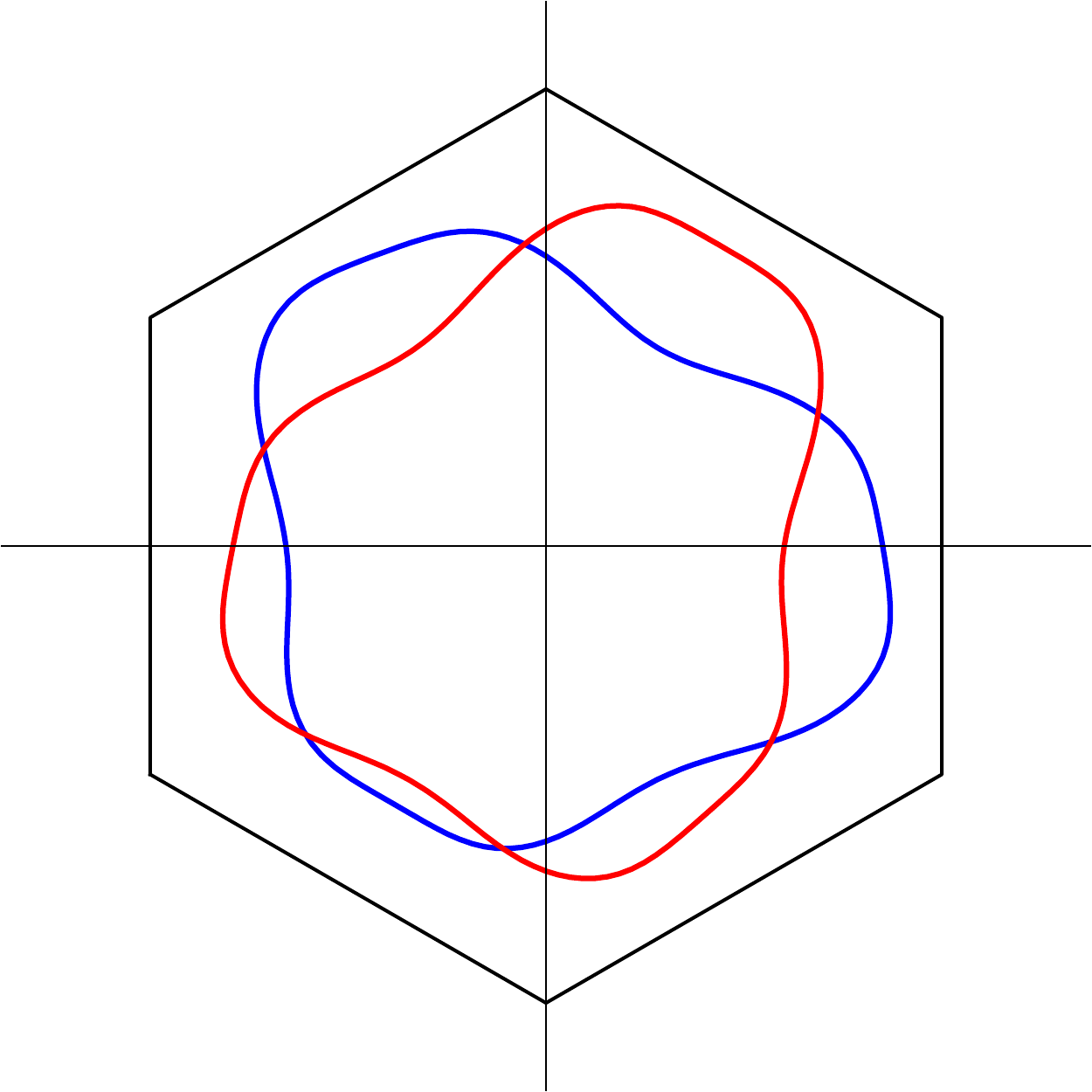}}
\subfigure[]{\includegraphics[width= 0.15 \textwidth]{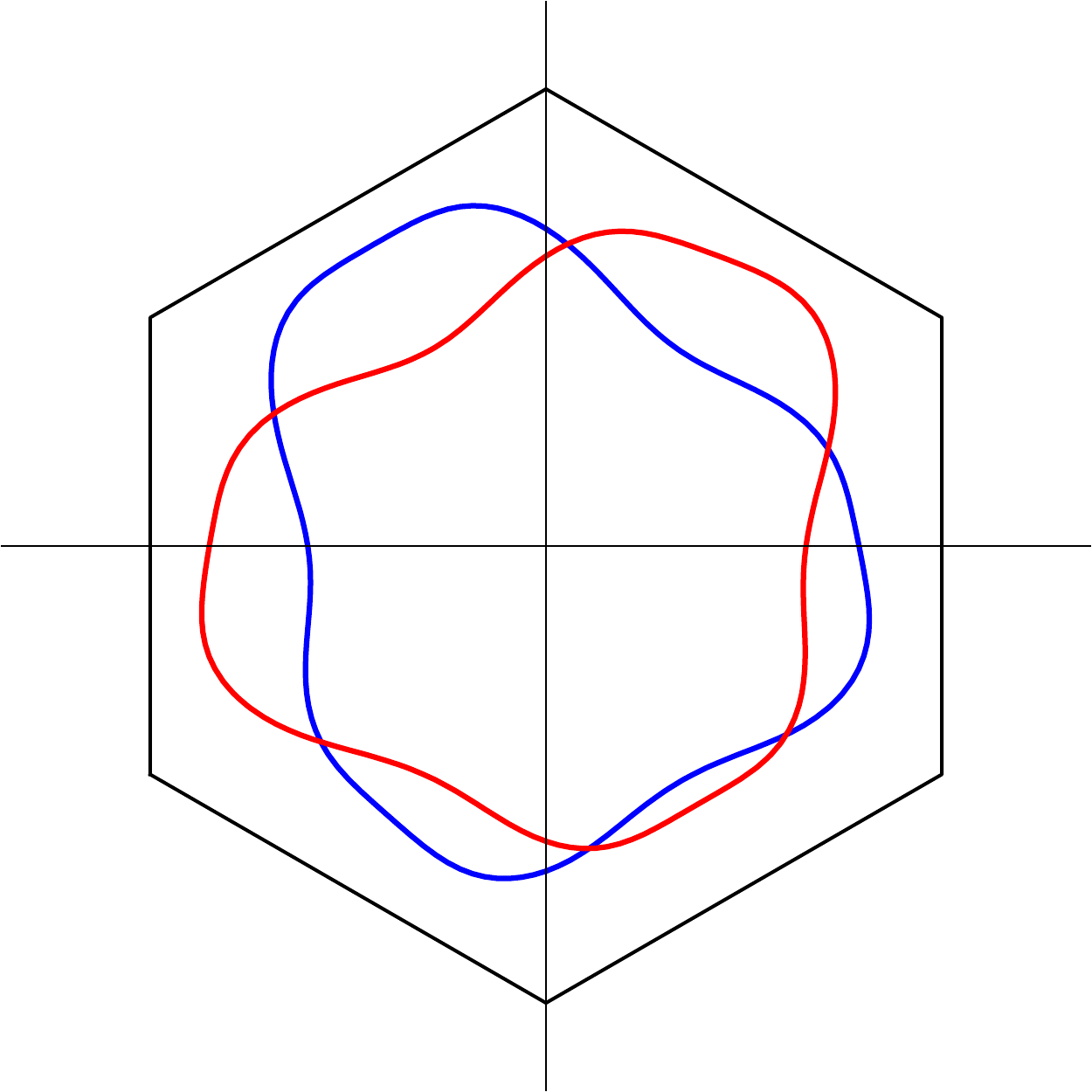}}
\subfigure[]{\includegraphics[width= 0.15 \textwidth]{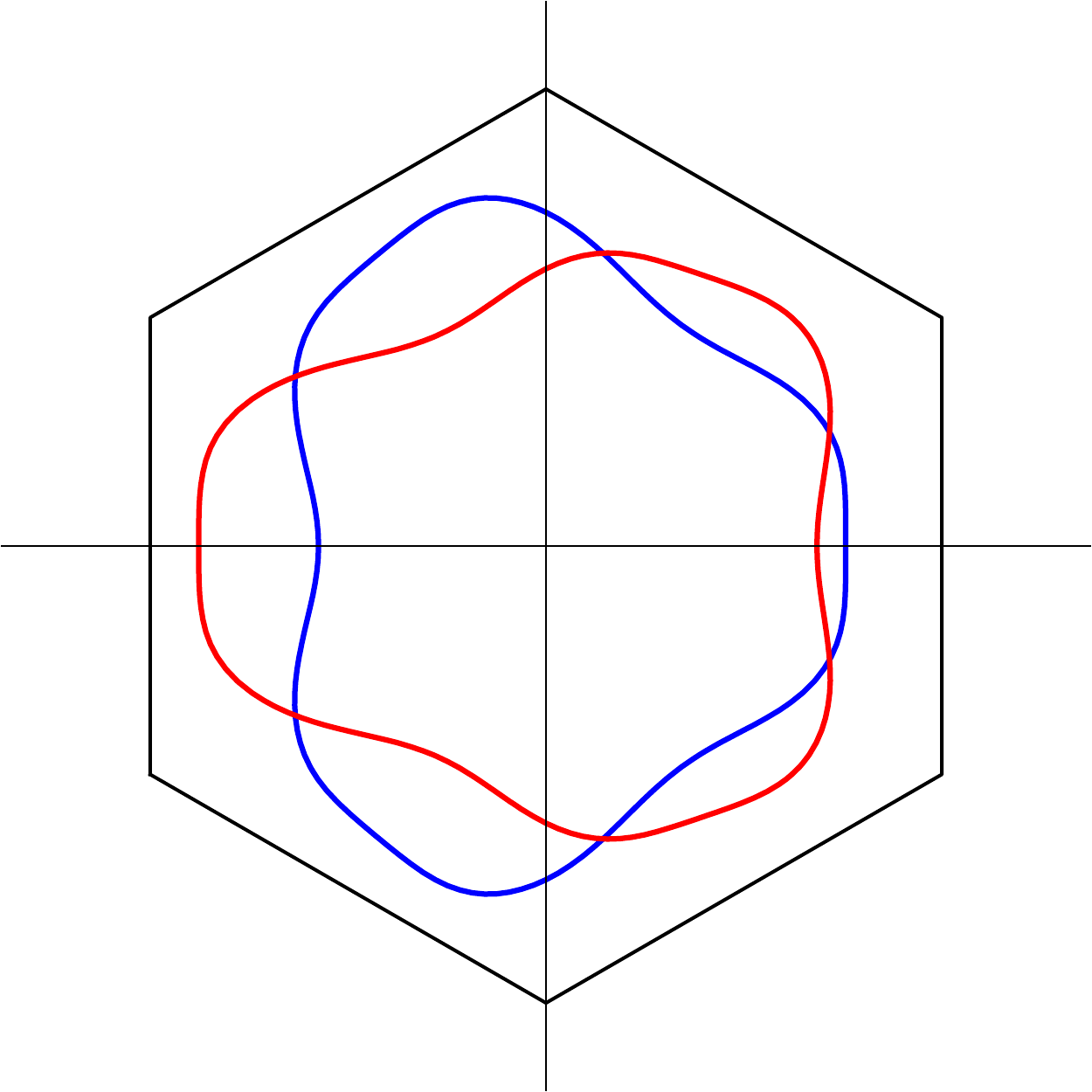}}
\subfigure[]{\includegraphics[width= 0.15 \textwidth]{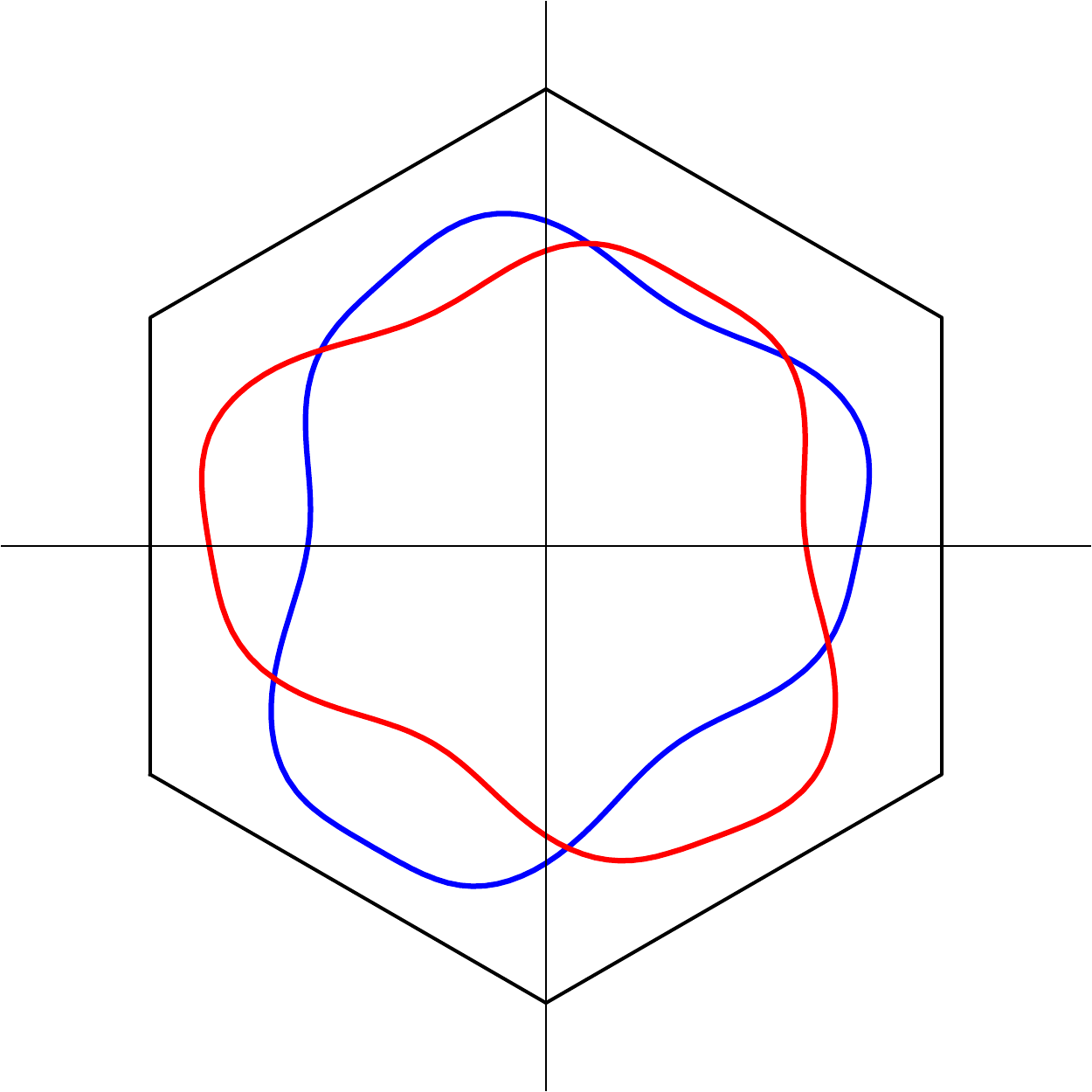}}
\subfigure[]{\includegraphics[width= 0.15 \textwidth]{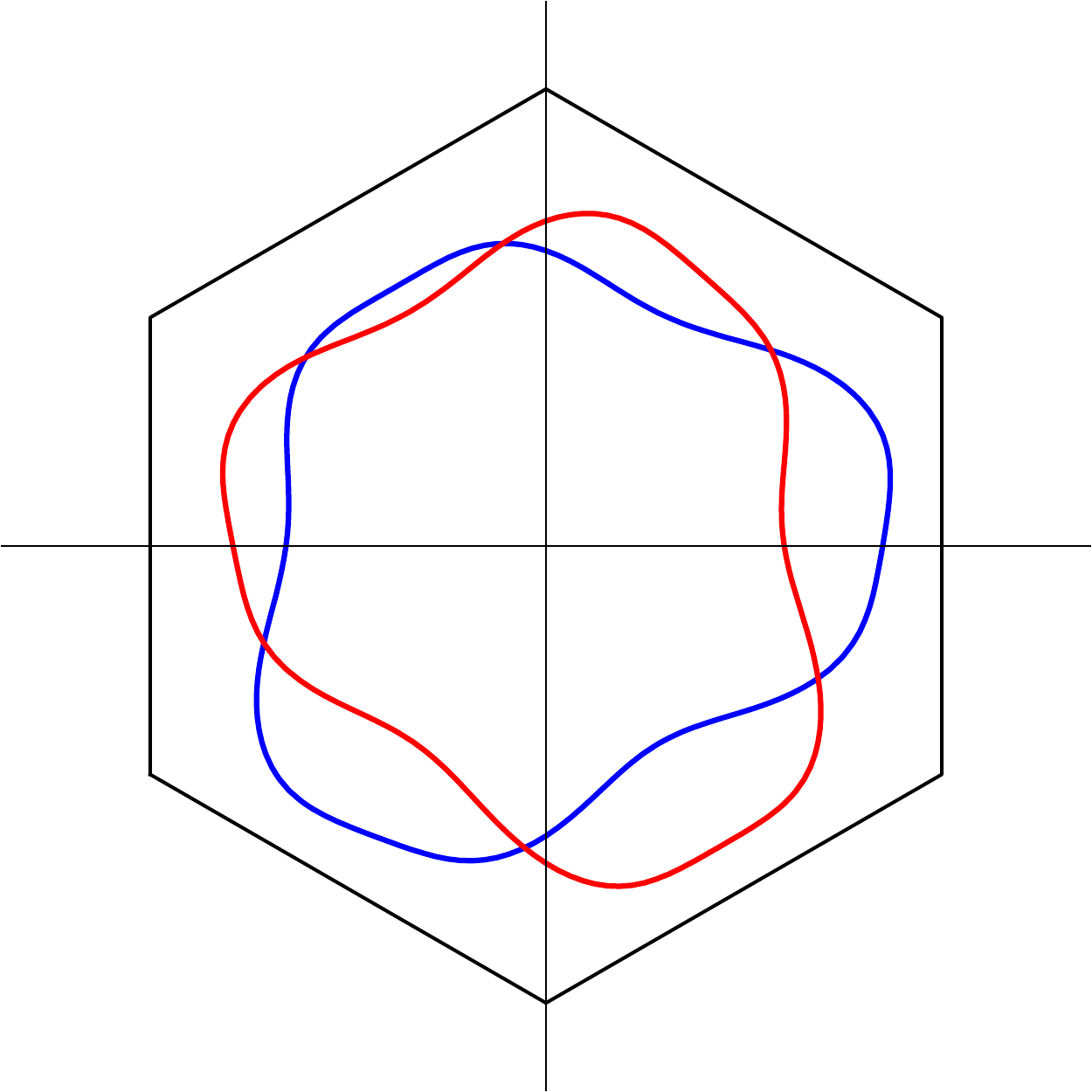}}
\caption{\label{fig:fermi-sur}
Fermi surfaces in the valley-polarized nematic state arising from the flat bands of TBG: The parameters are the same as those in Fig.~\ref{fig:bands}, except for $\alpha$, which here assumes the values $ n \,\pi/3$, with $n \in [0,5]$. Panels (a) to (f) correspond to $n=0$ to $n=5$, respectively, indicating the six different domains that minimize the free energy. In panel (a), the Fermi surfaces in the absence of valley-polarized nematic order are shown by the dashed lines. Red and blue lines refer to the two different valleys. 
}
\end{figure*}

To investigate how the valley-polarized nematic order parameter impacts
the electronic excitations of TBG, we use the six-band model of Ref. \cite{Po2019}. This model, which has valley $U(1)$ symmetry, is described in terms
of the electronic operator 
\begin{align}
\Psi_{a}^{\dagger}\left(\mathbf{k}\right)=(p_{a,{\bf k}z}^{\dagger},p_{a,{\bf k}+}^{\dagger},p_{a,{\bf k}-}^{\dagger},s_{1a,{\bf k}}^{\dagger},s_{2a,{\bf k}}^{\dagger},s_{3a,{\bf k}}^{\dagger})
\end{align}
for valley $a$, which contains the $p$-orbitals ($p_{z}$, $p_{+}$, $p_{-}$) living
on the sites of the triangular moir\'e superlattice, and $s$-orbitals
($s_{1}$, $s_{2}$, $s_{3}$) living on the sites of the related
Kagome lattice. The non-interacting Hamiltonian is given by
\begin{align}
\mathcal{H}_{0}=\sum_{\mathbf{k}}\left(\begin{array}{cc}
\Psi_{+}^{\dagger} & \Psi_{-}^{\dagger}\end{array}\right)\left(\begin{array}{cc}
H_{\mathbf{k}} & 0\\
0 & U_{C_{2z}}^{\dagger} H_{\mathbf{k}}\, U_{C_{2z}}
\end{array}\right)\left(\begin{array}{c}
\Psi_{+}\\
\Psi_{-}
\end{array}\right),
\label{eq:H0}
\end{align}
where the $6\times6$ matrices $H_{\mathbf{k}}$ and $U_{C_{2z}}$
are those defined in Refs. \cite{Po2019,Fernandes_Venderbos}. Generalizing the results of Ref. \cite{Fernandes_Venderbos}, the coupling to the valley-polarized
nematic order parameter $\Phi$ can be conveniently parametrized in
the $\left(p_{+}, \,p_{-}\right)$ subspace as
\begin{align}
\mathcal{H}_{\Phi}=\sum_{\mathbf{k}}\left(\begin{array}{cc}
\Psi_{+}^{\dagger} & \Psi_{-}^{\dagger}\end{array}\right)\left(\begin{array}{cc}
H_{\Phi} & 0\\
0 & -H_{\Phi}
\end{array}\right)\left(\begin{array}{c}
\Psi_{+}\\
\Psi_{-}
\end{array}\right),
\end{align}
with the block-diagonal matrix $H_{\Phi}=\left(0_{\mathbf{1}},\delta H_{\Phi},0_{\mathbf{3}}\right)$,
where
\begin{align}
\delta H_{\Phi}=\left(\begin{array}{cc}
0 & \Phi_0\,e^{-i \, \alpha}\\
\Phi_0\,e^{i\, \alpha} & 0
\end{array}\right).
\end{align}

In Fig.~\ref{fig:bands}, we show the electronic structure of the moir\'e flat bands
in the normal state (a) and in the valley-polarized nematic state (b) parametrized
by $\Phi_0=0.01 \,t_{\kappa}$ and $\alpha=0$, where $t_{\kappa}=27\ \mathrm{meV}$
is a hopping parameter of $H_{\mathbf{k}}$ \cite{Po2019}. The high-symmetry points
$\Gamma$, $K$, and $M$ all refer to the moir\'e Brillouin zone. The
main effect of the nematic valley-polarized order on the flat bands
is to lift the valley-degeneracy along high-symmetry directions. Although
$C_{2z}$ and $\mathcal{T}$ symmetries are broken, the combined symmetry
$C_{2z}\mathcal{T}$ remains intact in the valley-polarized nematic
phase. As a result, the Dirac cones of the non-interacting band structure
are not gapped, but instead move away from the $K$ points, similarly
to the case of standard (i.e. non-polarized) nematic order. We also
note that the Van Hove singularity at the $M$ point is also altered
by valley-polarized nematicity.

The Fermi surfaces corresponding to each of the six nematic valley-polarized
domains, described by $\alpha=n\, \pi/3$ with $n=0,1,\ldots,5$, are
shown in Fig.~\ref{fig:fermi-sur}. The Fermi surface of the normal state is also shown
in Fig.~\ref{fig:fermi-sur}(a) for comparison (dashed lines). In the ordered state, the Fermi surfaces arising from different valleys are distorted in different
ways, resulting in a less symmetric Fermi surface as compared with the
previously studied case of standard (i.e. non-polarized) nematicity.
While the Fermi surface is no longer invariant under out-of-plane
two-fold or three-fold rotations, it remains invariant under a two-fold
rotation with respect to an in-plane axis. Moreover, the Fermi surfaces
from different valleys continue to cross even in the presence
of valley-polarized nematic order.

\section{Pseudo-Goldstone modes in the valley-polarized nematic phase at zero temperature}

\label{secmodel}

In the previous section, we studied the general properties of valley-polarized nematic order in TBG. We now proceed to investigate the unique properties of the valley-polarized nematic state at $T=0$ in a metallic system, which stem from the emergence of a pseudo-Goldstone mode. As a first step, we extend the free energy in Eq. (\ref{eq:F6}) to a proper action. To simplify the notation, we introduce the complex valley-polarized nematic order parameter $\Phi = \Phi_1 - i \,\Phi_2 = \Phi_0 \,e^{i\, \alpha}$. We obtain (see also Ref.~\cite{cenke})
\begin{align}
S=\frac{1}{2}\int d^{2}x\,d\tau & \left[\frac{1}{c^{2}}\,|\partial_{\tau}\Phi|^{2}
+|\partial_{\mathbf{x}}\Phi|^{2}+r\,|\Phi|^{2}\right. \nn
 & \quad \left.+ \,u\,|\Phi|^{4}-\lambda\left(\Phi^{6}+{{\Phi}^{*}}^{6}\right)\right].
 \label{actionboson0}
\end{align}
Here, $\mathbf{x}$ denotes the position vector, $\tau$ denotes the imaginary time, and $c$ denotes the bosonic velocity. The quadratic coefficient $r$ tunes the system towards a putative quantum critical point (QCP) at $r=r_{c}$, and the quartic coefficient $u>0$. Because of the anisotropic $\lambda$-term, the action corresponds to a six-state clock model. As explained in Sec.~\ref{sec_phenomenology}, at finite temperatures, the behaviour of this model is the same as that of the two-dimensional (2D) six-state clock model. This model is known \cite{Jose1977} to first undergo a Kosterlitz-Thouless transition towards a state where the phase $\alpha$ has quasi-long-range order (like in the 2D XY model), which is then followed by another Kosterlitz-Thouless transition towards a state where $\alpha$ acquires a long-range order, pointing along one of the six directions that minimize the sixth-order term.

At $T=0$, near a valley-polarized nematic QCP, the bosonic model in Eq.~(\ref{actionboson0}) maps onto the three-dimensional (3D) six-state clock model \cite{Sudbo2003,Sandvik2021}. One of the peculiarities of this well-studied model is that the $\lambda$-term is a \emph{dangerously irrelevant perturbation} \cite{Oshikawa2000,Sandvik2007,Okubo2015,Leonard2015,Podolski2016,Sandvik2020}. Indeed, the scaling dimension $y$ associated with the $\lambda$ coefficient is negative; while an $\epsilon$-expansion around the upper critical dimension $d_{c}=4$ gives $y=-2-\epsilon$ \cite{Oshikawa2000}, recent Monte Carlo simulations report $y\approx-2.55$ for the classical 3D six-state clock model \cite{Okubo2015,Sandvik2020}. 

To understand what happens inside the ordered state, we use the parametrization $\Phi=\left|\Phi_{0}\right|e^{i\,\alpha}$, with fixed
$\left|\Phi_{0}\right|$, and consider the action for the phase variable $\alpha$ only, as shown below:
\begin{align}
S_{\alpha} 
& =\frac{1}{2}\int d^{2}x\,d\tau
\Bigg[\rho_{\tau}\,|\partial_{\tau}\alpha|^{2}
+\rho_{x}\,|\partial_{\mathbf{x}}\alpha|^{2}
\nn
& \hspace{2.5 cm }
-2\,\lambda\left|\Phi_{0}\right|^{6}\,\cos(6\alpha)\Bigg]\,.
\end{align}
Here, $\rho_{x}$ and $\rho_{\tau}$ are generalized stiffness coefficients.
Expanding around one of the minima of the last term (let us call it $\alpha_{0}$) gives
\begin{align}
S_{\alpha} & =\frac{1}{2}\int d^{2}x\,d\tau\Bigg[\rho_{\tau}
|\partial_{\tau}\,\tilde{\alpha}|^{2}
+\rho_{x}\, |\partial_{\mathbf{x}}\tilde{\alpha}|^{2}
\nn
& \hspace{2.5 cm } +36\left|\lambda\right|
\left|\Phi_{0}\right|^{6}\tilde{\alpha}^{2}\Bigg]\,,
\end{align}
where a constant term is dropped, and $\tilde{\alpha}\equiv\alpha-\alpha_{0}$.
It is clear that the $\lambda$-term, regardless of its sign, introduces a mass for the phase variable. Thus, while the $\lambda$-term is irrelevant at the critical point, which is described by the XY fixed point, it is relevant inside the ordered phase, which is described by a $Z_{6}$ fixed point, rather than the Nambu-Goldstone fixed point (that characterizes the ordered phase of the 3D XY model) \cite{Oshikawa2000,Sandvik2020,Sandvik2021}. 

Importantly, due to the existence of this dangerously irrelevant perturbation, there are two correlation lengths in the ordered state \cite{Sandvik2007,Okubo2015,Leonard2015,Podolski2016,Sandvik2021}: $\xi$ associated with the usual amplitude fluctuations of $\Phi$; and $\xi'$ associated with the crossover from continuous to discrete symmetry-breaking of $\alpha$. Although both diverge at the critical point, they do so with different exponents $\nu$ and $\nu'$, respectively. Because $\nu'>\nu$, there is a wide range of length scales (and energies, in the $T=0$ case) for which the ordered phase behaves as if it were an XY ordered phase. In Monte Carlo simulations, this is signalled by the emergence of a nearly-isotropic order parameter distribution \cite{Sandvik2007}.
More broadly, this property is expected to be manifested as a small gap in the spectrum of phase fluctuations, characteristic of a \emph{pseudo-Goldstone mode} \cite{Burgess2000}.

For simplicity of notation, in the remainder of this paper, we rescale $\left(\tau,\mathbf{x}\right)$ to absorb the stiffness coefficients. Moreover, we set $\lambda>0$ and choose $\alpha_{0}=0$, such that $\tilde{\alpha}=\alpha$. Defining $m^{2}\equiv36\left|\lambda\right|\left|\Phi_{0}\right|^{6}$,
and taking the Fourier transform, the phase action becomes
\begin{align}
S_{\alpha}=\frac{1}{2}\int_{q}\alpha(-q)\left(\omega_{n}^{2}+\mathbf{q}^{2}+m^{2}\right)\alpha(q)\,,
\label{S_phase}
\end{align}
where $q=\left(\omega_{n},\mathbf{q}\right)$, $\omega_{n}$ is the bosonic Matsubara frequency, and $\mathbf{q}$ is the momentum. Here,
we also introduced the notation $\int_{q}=
T \sum \limits_{\omega_n}\int\frac{d^{2}\mathbf{q}}{(2\pi)^{2}}$. At $T=0$, $T \sum \limits_{\omega_n} \rightarrow \int 
\frac{d\omega_n} {2\pi}$; although the subscript $n$ is not necessary, since $\omega_n$ is a continuous variable, we will keep it to distinguish it from the real-axis frequency.

Having defined the free bosonic action, we now consider the electronic degrees of freedom. While our work is motivated by the properties of TBG, in this section we choose a simple, generic band dispersion to shed light on the general properties of the $T=0$ valley-polarized nematic state.
As we will argue later, this formalism also allows us to discuss the case of a spin-polarized nematic state. The free fermionic action is given by:
\begin{align}
S_{f}=\int_{k}\sum\limits _{a=1,2}\psi_{a}^{\dagger}(k)
\left[i\,\nu_{n}
+\varepsilon_{a} (\mathbf{k} )\right]\psi_{a}(k)\,,
\label{S_f}
\end{align}
where $k =\left( \nu_n, \mathbf k \right)$, $a$ is the valley index, and $\nu_{n}$ is the fermionic Matsubara frequency. The electronic dispersion $\varepsilon_{a}\left(\mathbf{k}\right)$ of valley $a$ could, in principle, be derived from the tight-binding model of Eq. (\ref{eq:H0}); for our purposes, however, we keep it generic. In this single-band version of the model, the valley-polarized nematic order parameter couples to the fermionic degrees of freedom as described by the action \cite{cenke}
\begin{align}
\label{S_bf}
 S_{bf} & =\gamma_{0}\int_{k,q}\sum\limits _{a=1,2}(-1)^{a+1} \,
 \psi_{a}^{\dagger}(k+q)\,\psi_{a}(k) \nn
 & \times\left[\frac{\Phi(q)+\Phi^{*}(q)}{2}\cos(2\theta_{k})-\frac{\Phi(q)-\Phi^{*}(q)}{2\,i}\sin(2\theta_{k})\right].
\end{align}
Here, $\gamma_{0}$ is a coupling constant, and $\tan\theta_{k}=k_{y}/k_{x}$.
Writing $\Phi=\left|\Phi_{0}\right|e^{i\,\alpha}$, we obtain the coupling between the phase variable and the electronic operators inside the valley-polarized nematic state with constant $\left|\Phi_{0}\right|$. As before, we set $\alpha_{0}=0$, and expand around the minimum, to obtain
\begin{align}
\label{S_alphaf}
S_{\alpha f} & =\gamma\int_{k,q}\sum\limits _{a=1,2}(-1)^{a+1} \,
\psi_{a}^{\dagger}(k+q)\,\psi_{a}(k) \nn
 & \hspace{1 cm}
 \times\left[\cos(2\theta_{k})\left(2\pi\right)^{3}\delta^{3}(k-q)
-  \alpha(q)\,\sin(2\theta_{k})\right] . 
\end{align}
where $\gamma\equiv\gamma_{0}\left|\Phi_{0}\right|$. The first term
in the last line shows that long-range order induces opposite nematic distortions in the Fermi surfaces with opposite valley quantum numbers. The second term shows that the phase mode couples to the charge density directly via a Yukawa-like coupling. As discussed in Ref.~\cite{Ashvin14}, this is an allowed coupling when the generator of the broken symmetry does not commute with the momentum operator.

\section{Non-Fermi liquid to Fermi liquid crossover}

\label{secnfl}

\subsection{The patch model}

Our goal is to derive the properties of the electronic degrees of
freedom in the valley-polarized nematic ordered phase, which requires the computation of the electronic self-energy. To do that in a controlled manner, we employ the patch method discussed in Ref.~\cite{max-subir,Lee-Dalid,ips-lee,ips-uv-ir,ips-fflo,ips-nfl-u1}.
This relies on the fact that fermions from different patches of a Fermi surface interact with a massless order parameter with largely disjoint sets of momenta, and that the inter-patch coupling is small in the low-energy limit, unless the tangent vectors at the patches are locally parallel or anti-parallel. Thus, the advantage of this emergent locality in momentum space is that we can now decompose the full theory into a sum of two-patch theories, where each two-patch theory describes electronic excitations near two antipodal points,
interacting with the order parameter boson with momentum along the local tangent.
This formalism has been successfully used in computing the universal properties and scalings for various NFL systems, such as the Ising-nematic QCP \cite{max-subir,Lee-Dalid,ips-lee,ips-uv-ir,ips-subir,ips-sound}, the Fulde-Ferrell-Larkin-Ovchinnikov (FFLO) QCP \cite{ips-fflo}, and a critical Fermi surface interacting with transverse gauge field(s) \cite{ips-nfl-u1}.
The only scenario that breaks this locality in momentum space is the presence of short-ranged four-fermion interactions in the pairing channel \cite{max-cooper,ips-sc}.

\begin{figure}
\centering \includegraphics[width=0.3\textwidth]{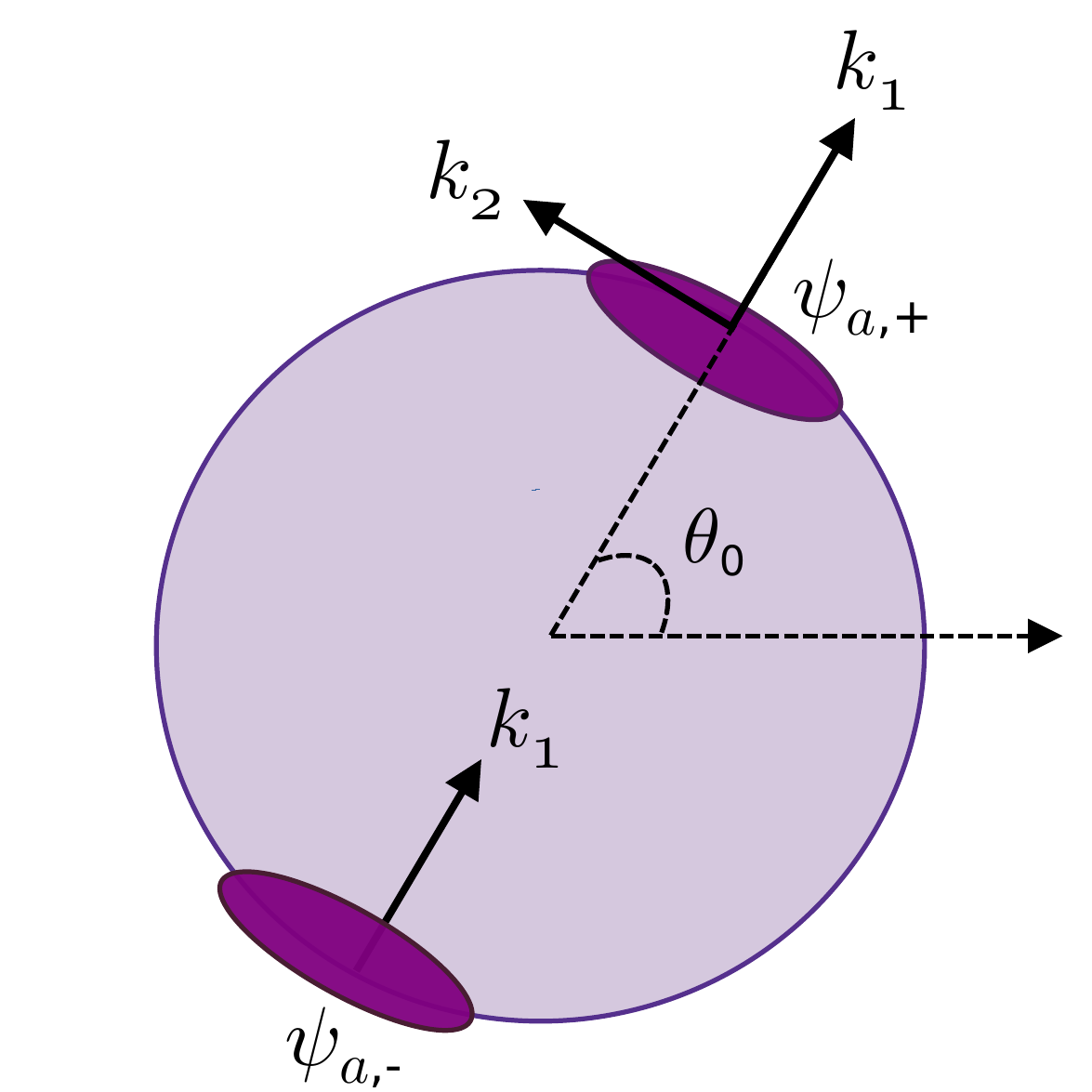} 
\caption{\label{figpatch}
Illustration of the patch model: $\psi_{a,+}$ denotes
the fermions located at the upper light purple patch, centered at an angle $\theta=\theta_{0}$ with respect to the global coordinate system for a circular Fermi surface of valley quantum number $a$ (denoted by the dark purple ring). $\psi_{a,-}$ denotes the fermions in the lower light purple patch, centred at the antipodal point $\theta=\pi+\theta_{0}$, whose tangential momenta are parallel to those at $\theta_{0}$. Although we show here the patch construction for a circular Fermi surface for the sake of simplicity, this can be applied to any Fermi surface of
a generic shape, as long as it is locally convex at each point.}
\end{figure}

For our case of the valley-polarized nematic order parameter, we consider two antipodal patches on a simplified Fermi surface, which is locally convex at each point. The  antipodal patches feature opposite Fermi velocities, and couple with the bosonic field \cite{Lee-Dalid,ips-lee,ips-uv-ir,ips-fflo}.
Here, we choose a patch centred at $\theta_{k}=\theta_{0}$, and construct our coordinate system with its origin at $\theta_{0}$. As explained above, we must also include the fermions at the antipodal patch with $\theta_{k}=\pi+\theta_{0}$.
We denote the fermions living in the two antipodal patches as $\psi_{+}$ and $\psi_{-}$, as illustrated in Fig. \ref{figpatch}. We note that
the coupling constant remains the same for the fermions in the two antipodal points.

Expanding the spectrum around the Fermi surface patches up to an effective parabolic dispersion, and using Eqs.~(\ref{S_phase}),
(\ref{S_f}), and (\ref{S_alphaf}), we thus obtain the effective
field theory in the patch construction as
\begin{align}
S_{\text{tot}} & = S_{f} + S_{\alpha}  + S_{\alpha f}\,,
\text{where}\nonumber \\
S_{f} & =\int_{k}\sum\limits _{\substack{s=\pm \\a=1,2}}
\psi_{a,s}^{\dagger}(k)
\Big(
i\,\nu_{n}+s\,k_{1}+\frac{k_{2}^{2}}{2\,k_{F}} \Big)
 \psi_{a,s}(k)\,,\nonumber \\
 S_{\alpha} 
 & =\frac{1}{2}\int_{q} \alpha(-q)
\left(\omega_{n}^{2}+q_{1}^{2}
+ q_{2}^{2}+m^{2}\right)\alpha(q)\,,
\nonumber \\
 S_{\alpha f} &=\sum\limits_{\substack{s=\pm\\a=1,2}}  (-1)^{a} 
\int_{k,q} 
\psi_{a,s}^{\dagger}(k+q)\,
\Big [
\gamma\sin(2\theta_{0})\, \alpha(q)
\nonumber \\ & \hspace{3 cm} 
  - \gamma\cos(2\theta_{0})
 \Big ] \,\psi_{a,s}(k)\,.
\end{align}
Here, for simplicity, we have assumed that the Fermi surface is convex, and has the same shape for both the valley quantum numbers. We will discuss the impact of these approximations later in this section. Note that the fermionic momenta are expanded about the Fermi momentum $k_{F}$ at the origin of the coordinate system of that patch. In our notation, shown in Fig. \ref{figpatch}, $k_{1}$ is directed along the local Fermi momentum, whereas $k_{2}$ is perpendicular to it (or tangential to the Fermi surface). Note that the local curvature of the Fermi surface is given by
$1/ k_{F}$. Furthermore, $\psi_{a,+}$ ($\psi_{a,-}$) is the right-moving (left-moving) fermion with valley index
$a$, whose Fermi velocity along the $k_{1}$ direction is positive (negative).

Following the patch approach used in Refs.~\cite{Lee-Dalid,ips-lee,ips-uv-ir,ips-nfl-u1},
we rewrite the fermionic fields in terms of the two-component spinor $\Psi$, where
\begin{align}
\Psi^{T}(k) & =\left(\psi_{1,+}(k)\quad\psi_{2,+}(k)\quad\psi_{1,-}^{\dagger}(-k)\quad\psi_{2,-}^{\dagger}(-k)\right),\nonumber \\
\bar{\Psi}(k) & =\Psi^{\dagger}(k)\,\sigma_{2}\otimes\tau_{0}\,.
\end{align}
Here, $\sigma_{i}$ (with $i=1,2, 3$) denotes the $i^{\rm{th}}$ Pauli matrix acting on the patch space (consisting of the two antipodal patches), whereas $\tau_{i}$ is the $i^{\rm{th}}$ Pauli matrix acting on valley space (not to be confused with imaginary time $\tau$, which has no subscript). We use the symbols $\sigma_{0}$ and $\tau_{0}$ to denote the corresponding $2\times 2$ identity matrices.
In this notation, the full patch action
$S_{\text{tot}}  = S_{f} + S_{\alpha}  + S_{\alpha f}$ consists of
\begin{widetext}
\begin{align}
S_{f} & =\int_{k}\bar{\Psi}^{\dagger}(k)\left[
i\left(  \sigma_2 \,\nu_n
+\sigma_{1}\,\delta_{k}\right)
\otimes\tau_{0} \right]\Psi(k)\,,\quad
S_{\alpha}=\frac{1}{2}\int_{q}\alpha(-q)
\left(\omega_n^{2}+q_{1}^{2}+q_{2}^{2}+m^{2}\right)\alpha(q)\,,\nonumber \\
S_{\alpha f} & =\gamma\int_{k,q}\bar{\Psi}(k+q)\left[
\left(2\pi\right)^{3}
\delta^{3}(k-q)\,\cos(2\theta_0)\,\sigma_{2}
- i\,\sin(2\theta_0)\,\alpha(q)\,\sigma_{1}\right]
\otimes\tau_{3} \, \Psi(k)\,,\quad\delta_{k}=k_{1}+\frac{k_{2}^{2}}{2\,k_{F}}\,.
\label{model}
\end{align}
\end{widetext}
For convenience, we have included the valley-dependent Fermi surface distortion $\gamma \cos(2\theta_{0})$
in the interaction action. The form of $S_{f}$ is such that it appears as if the fermionic energy disperses only in one effective direction near the Fermi surface. Hence, according to the formulation of the patch model in Refs.~\cite{Lee-Dalid,ips-lee,ips-uv-ir,ips-fflo}, the $(2+1)$-dimensional fermions can be viewed as if they were a $(1+1)$-dimensional ``Dirac'' fermion, with the momentum along the Fermi surface interpreted as a continuous flavor.

From Eq.~(\ref{model}), the bare fermionic propagator can be readily obtained as
\begin{align}
G_{0}(k)=-i\,\frac{\sigma_{2}\,\nu_{n} + \sigma_{1}\,\delta_{k}}
{\nu_{n}^{2}+\delta_{k}^{2}}\otimes\tau_{0}\,.
\label{fermprop}
\end{align}
We note that the strength of the coupling constant between
the bosons and the fermions, given by $\gamma\,\sin(2\theta_k)$, depends on the value of $\theta_k $. For the patch centered at $\theta_k =\theta_{0}$,
the leading order term from the loop integrals can be well-estimated by assuming $\theta=\theta_{0}$ for the entire patch, as long as $\sin(2\theta_{0})\neq0$. However, for $\sin(2\theta_{0})=0$, we need to go beyond the leading order terms (which are zero), while performing the loop integrals. The patches centered around $\theta_k =\theta_0 $, with $\sin(2\theta_{0})\sim0$, are the so-called ``cold spots''; we will refer to the other patches as belonging to the ``hot regions'' of the Fermi surface.

\subsection{Electronic self-energy}

We first compute the one-loop bosonic self-energy $\Pi_{1}$, which takes the form:
\begin{widetext}
\begin{align}
\Pi_{1}(q) & =-\left(i\,\gamma\right)^{2}\int\frac{d^{3}k}{(2\,\pi)^{3}}\left[\sin^{2}(2\theta_{0})
+\frac{4\,k_{2}^{2}\cos(4\theta_{0})}{k_{F}^{2}}+\frac{2\,k_{2}\sin(4\theta_{0})}{k_{F}}\right]\text{Tr}\left[\sigma_{1}\,G_{0}(k+q)\,\sigma_{1}G_{0}(k)\right]\nonumber \\
 & =-\frac{\gamma^{2}\sin^{2}(2\theta_{0})\,k_{F}\,|\omega_{n}|}{\pi\,|q_{2}|}+\frac{2\,\gamma^{2}\sin(4\theta_{0})\,k_{F}\,\delta_{q}\,|\omega_{n}|}{\pi\,q_{2}\,|q_{2}|}+\frac{4\,\gamma^{2}\cos(4\theta_{0})\,k_{F}\,|\omega_{n}|\left[\pi\left(q_{0}^{2}-\delta_{q}^{2}\right)-2|q_{0}|\,|q_{2}|\right]}{\pi^{2}\,|q_{2}|^{3}}\,.
\end{align}
\end{widetext}
This result is obtained by considering a patch centered
around $\theta_k =\theta_{0}$ and then expanding $\sin^{2}(2\theta_{0}+2\,k_{2}/k_{F})$
in inverse powers of $k_{F}$. In the limits $\frac{\left|\omega_{n}\right|}{|q_{2}|}\ll1$,
$k_{F}\gg|\mathbf{q}|$, and $|\mathbf{q}|\rightarrow\mathbf{0}$,
we have, to leading order
\begin{align}
\Pi_{1}(q)\Big\vert_{\text{hr}}
=-\frac{\left|\omega_{n}\right|}{|q_{2}|}\frac{\gamma^{2}\sin^{2}(2\theta_{0})\,k_{F}}{\pi},
\end{align}
as long as $\sin(2\theta_{0})\neq0$ (i.e., in the hot regions). For the cold spots, the leading-order term is given by 
\begin{align}
\Pi_{1}(q)\Big\vert_{\text{cs}}=-\frac{8\,\gamma^{2}\cos(4\theta_{0})\,k_{F}\,}{\pi^{2}}\frac{\omega_{n}^{2}}{q_{2}^{2}},
\end{align}
Here, the subscript ``hr'' (``cs'') denotes hot regions (cold spots).
A similar result was previously obtained in Refs.~\cite{Oganesyan01,Garst09} using a different approach, and for the case of an XY nematic order parameter (see also \cite{Carvalho2019}). Therefore, we conclude that the pseudo-Goldstone mode in the valley-polarized nematic phase is overdamped in the hot regions. 

\begin{figure*}
\centering 
\includegraphics[width=0.60 \textwidth]{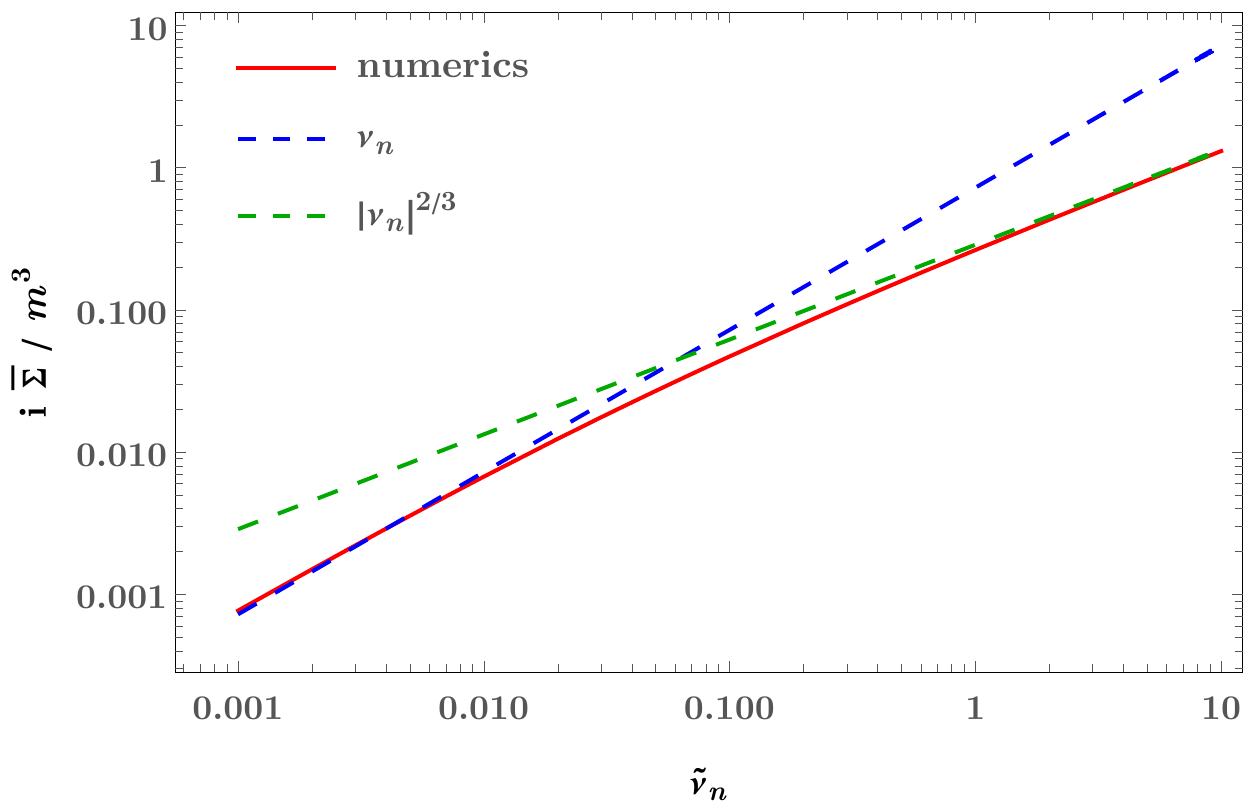}
\caption{\label{figsigma} Fermionic self-energy $i\,\bar{\Sigma}(\nu_{n})/m^3$
as a function of the scaled Matsubara frequency $\tilde \nu_{n} = \nu_n/\Omega^* $, obtained from the numerical integration of Eq.~(\ref{sigma_numerics}) by setting $m = 0.1$ and $ k_F = 100$.
The dashed lines correspond to the frequency dependencies obtained
from the asymptotic results in Eq.~(\ref{sigma_NFL}) [ i.e., $i\,\bar{\Sigma}\left(\nu_{n}\right)\sim\left|\nu_{n}\right|^{2/3}$] and in Eq.~(\ref{sigma_FL}) [i.e., $i\,\bar{\Sigma}\left(\nu_{n}\right)\sim\nu_{n}$].}
\end{figure*}

We can now define the dressed bosonic propagator, which includes the one-loop bosonic self-energy, as
\begin{align}
D_{1}(q)=\frac{1}{q^{2}+m^{2}-\Pi_{1}(q)}\,.
\label{eqbosprop}
\end{align}
The one-loop fermionic self-energy $\Sigma_{1}(k)$ can then be expressed in terms of $\tilde{\Sigma}$, defined as
\begin{align}
 & \tilde{\Sigma}(k)\equiv\Sigma_{1}(k)-\gamma\,\cos(2\,\theta_{0})\,\sigma_{2}\otimes\tau_{3}\nonumber \\
 & =-\gamma^{2}\sin^{2}(2\theta_{0})\int_{q}\left(\sigma_{1}\otimes\tau_{3}\right)G_{0}(k+q)\left(\sigma_{1}\otimes\tau_{3}\right)D_{1}(-q)\,,
 \label{eqselfen}
\end{align}
where we use the notation $q=(\omega_{n'}, \mathbf q)$.
In order to be able to perform the integrals, we will neglect the $q_{1}^{2}$ and $\omega_{n'}^{2}$ contributions in the bosonic propagator, which are anyway irrelevant
in the RG sense \cite{Lee-Dalid,ips-lee}.
This is justified because the contributions to the integral
are dominated by $q_{1}\sim \nu_{n}$, $ \omega_{n'} \sim \nu_{n}$, and $q_{2}\sim|\nu_{n}|^{1/3}$,
and we are interested in the small $|\nu_{n}|$ limit (where $\nu_{n}$ is the external fermionic Matsubara frequency). In the limit $m\rightarrow0$,
we can obtain analytical expressions for $\tilde{\Sigma}(k)$
as follows: 
\begin{align}
 & \tilde{\Sigma}(k)\Big\vert_{\text{hr},m\rightarrow0}\nonumber \\
 & =-\gamma^{2}\sin^{2}(2\theta_{0})\int_{q}\left(\sigma_{1}\otimes\tau_{3}\right)G_{0}(k+q)\left(\sigma_{1}\otimes\tau_{3}\right)D_{1}(-q)
\nonumber \\
 & =
 - \frac{i
 \left[\gamma\sin(2\theta_{0})\right]^{4/3}
 \text{sgn}(\nu_{n})\,|\nu_{n}|^{2/3}}{2\,\sqrt{3}\,\pi^{2/3}\,k_{F}^{1/3}}\,\sigma_{2}\otimes\tau_{0}\,,
\label{sigma_NFL}\\
 & \tilde{\Sigma}(k)\Big\vert_{\text{cs},m\rightarrow0}
\nn & =
- \frac{i\,\gamma^{3/2}\cos^{\frac{3}{4}}\left(4\theta_{0}\right)\text{sgn}(\nu_{n})\,|\nu_{n}|^{1/2}
 \,k_{2}^{2}
}
 {2^{1/4}\,\sqrt{\pi}\,k_{F}^{9/4}}
 \,\sigma_{2}\otimes\tau_{0} \,.
\end{align}
The one-loop corrected self-energy is then given by $G^{-1}(k) = G_0^{-1}(k) -\Sigma_1(k) $. The frequency dependence of $\tilde{\Sigma}$ in the hot regions, in the limit $m\rightarrow0$, corresponds to an NFL behaviour, since the fermionic lifetime has a sublinear dependence on frequency, implying the absence of well-defined quasiparticles. The same $|\nu_n|^{2/3}$ dependence on the frequency
was found in the case of an ideal XY nematic in Refs.~\cite{Oganesyan01,Garst2010}.

However, for the valley-polarized nematic state, $m$ is not zero in the ordered state, as it is proportional to the square root of the dangerously irrelevant variable $\lambda$ in the bosonic action. The limit of large $m$ is straightforward to obtain, and gives an FL-correction to the electronic Green's function, because
\begin{align}
& \tilde{\Sigma}(k)\Big\vert_{\text{hr},m\gg
\left[
\frac{3 \,\sqrt{3} \,\gamma ^2 \,k_F
\sin^2\left(2 \theta _0\right) \,\left| \nu _n\right| }
{2 \,\pi }
\right]^{1/3}
} 
 \nn &
= -
\frac{ \left(2+2^{2/3}\right) 
\gamma ^2 \sin ^2\left(2 \theta _0\right)}
{8 \,\pi\,  m}\,i\,\nu_n \,\sigma_{2}\otimes\tau_{0}\,.
\label{sigma_FL}
\end{align}
From Eqs.~\eqref{fermprop}, \eqref{eqbosprop}, and \eqref{eqselfen}, we find that the crossover from NFL to FL behaviour occurs when $m^2> -\Pi_1(q)$, i.e., $m^2>\frac{\left| \omega _{n'}\right| \,\gamma ^2 k_F \sin^2\left(2 \theta _0 \right)}
{\pi \, q_2}$ in the one-loop corrected bosonic propagator $D_1(q)$ inside the integral. In that situation,
the dominant contribution to the integral over $q_2$ comes from $q_2\sim m$.
On the other hand, considering the fermionic propagator contribution to the integrand, the dominant contribution comes from $ \omega_{n'} \sim \nu_n$ for the $ \omega_{n'}$-integral.
Hence, the relevant crossover scale for the fermionic frequency $\nu_n$ is approximately $\Omega^{*}  = \frac{m^3}
{\gamma ^2 \,k_F
\sin^2\left(2 \theta _0\right)} $.
Because $m^2 \sim \lambda$, it follows that $\Omega^* \sim \lambda^{3/2}$.

It is therefore expected that, for finite $m$, above the characteristic
energy scale $\Omega^{*}$,
the self-energy displays NFL
behaviour, captured by $\tilde{\Sigma}\sim i\,\text{sgn}(\nu_{n})\,|\nu_{n}|^{2/3}$.
For low enough energies, such that $\left|\nu_{n}\right|\ll\Omega^{*}$, the regular FL behaviour with $\tilde{\Sigma}\sim i \,\nu_{n}$ should be recovered. The crucial point is that, because $\Omega^{*}$ depends on the dangerously irrelevant coupling constant $\lambda$, it is expected to be a small energy
scale. This point will be discussed in more depth in the next section. To proceed, it is convenient to write the complete expression for
$\tilde{\Sigma} = \bar{\Sigma} \times \left( \sigma_{2}\otimes\tau_{0} \right )$ for the case of an arbitrary $m$:
\begin{align}
&  i\,\bar{\Sigma}(k)\Big\vert_{\text{hr}} 
\nn &  = 
-\int  {d \tilde \omega_{n'}} 
\frac{  m^3\,\text{sgn}\left( \tilde \nu _n+ \tilde \omega _{n'}\right)}
{4\, \pi ^2} 
\, \sum \limits_{j=1}^3
\frac{\zeta_j(\tilde \omega_{n'}) \ln \left (-\zeta_j(\tilde \omega_{n'})\right )}
{m^2+ 3 \,\zeta_j^2(\tilde \omega_{n'})}\,,
\label{sigma_numerics}
\end{align}
where $\tilde{\nu}_{n}\equiv\nu_{n}/\Omega^{*}$, $\tilde{\omega}_{n'}\equiv\omega_{n'}/\Omega^{*}$, and $\zeta_j $ is the $j^{\text{th}}$ root of the cubic-in-$q_2$ polynomial
$
\pi  \,q_2 \left( q_2^2+ m^2 \right) +
m^3 \,k_F \left| \tilde{\omega }_{n'}\right| 
$.

To confirm that indeed $\Omega^{*}$ is the energy scale associated with the crossover from NFL to FL behaviour, we have solved the integral in Eq.~(\ref{sigma_numerics}) numerically to obtain the self-energy for arbitrary $m$. As shown in Fig.~\ref{figsigma}, $\Omega^{*}$ separates the two asymptotes for the self-energy $\tilde{\Sigma}$:
(1) NFL, given by Eq. (\ref{sigma_NFL}), and present for $\nu_{n}\gg\Omega^{*}$;
(2) FL, given by Eq. (\ref{sigma_FL}), and present for $\nu_{n}\ll\Omega^{*}$.

As explained in the beginning of this section, here we have considered the simplified case of two identical convex Fermi surfaces for the two valleys. This not only makes the analytic calculations more tractable, but also allows us to extend the results for more general cases beyond TBG. This includes, for instance, the case where $a$ is not a valley quantum number, but a spin quantum number, which we will discuss in more detail in Sec. \ref{secend}.

Considering the Fermi surfaces for TBG obtained from the tight-binding model and shown in Fig.~\ref{fig:fermi-sur}, it is clear that they each have a lower three-fold (rather than continuous) rotational symmetry in the disordered state. One of the consequences is that the two patches in the patch model are no longer related by inversion, at least not within the same valley Fermi surface. Another consequence is that the Fermi surface can have points that are locally concave and not convex. The latter is an important assumption of the patch model construction of Ref.~\cite{Lee-Dalid,ips-nfl-u1}, which we have implemented here. The impact of these two effects on the self-energy behaviour at moderate frequencies is an interesting question beyond the scope of this work, which deserves further investigation. While we still expect an FL-to-NFL crossover, the particular frequency dependence of the self-energy, in the regime where the pseudo-Goldstone mode appears gapless, may be different from what has been discussed in this section.

\subsection{Hertz-Millis approach}

We note that the same general results obtained above also follow from the usual (but uncontrolled) Hertz-Millis approach~\cite{Wolfle_RMP}. It turns out that the action in Eq.~(\ref{S_alphaf}) is analogous to the widely studied case of a metallic Ising-nematic QCP, and hence the results are well known (see, for example, Refs.~\cite{Metzner03,max-subir,Hartnoll2014,Paul2017}). Linearizing the dispersion near the Fermi level, the one-loop bosonic self-energy is given by
\begin{widetext}
\begin{align}
 \bar{\Pi}_1(q) 
 = - \gamma^2 \,k_F
\int_{-\infty}^{\infty}\,\frac{d \nu_{n'} }{2\,\pi}
\int_{-\infty}^{\infty}\,\frac{dk_\perp}{2\,\pi}
\int_{0}^{2\pi}\,\frac{d\theta_k} {2\,\pi}  
\,
\frac{\sin^2 (2\theta_k)}  { \left( i \,\nu_{n'} - k_\perp\right )
\left[ i \left ( \nu_{n'} + \omega_n \right) -
\left \lbrace k_\perp + |\mathbf q | \cos(\theta_k - \theta_q) \right \rbrace \right ]}
\,.
\end{align}
\end{widetext}
A straightforward computation gives the final expression:
\begin{align}
\bar{\Pi}_1(q) \propto - \gamma ^2  \sin ^2\left(2 \theta _q\right) \,\frac{\left| \omega _n\right| } { | \mathbf q|}\,.
\end{align}
Thus, we obtain the dynamical critical exponent $z=3$ for the bosons [except at the cold spots, where the coupling constant $\sin ^2\left(2 \theta _q\right)$ vanishes]. This is the usual Hertz-Millis result for a bosonic QCP in a metal, whose ordered state has zero wavevector \cite{Wolfle_RMP}. Most importantly, it gives an NFL fermionic self-energy $\bar{\Sigma}_1 \propto i \, |\nu_n|^{2/3}$ if the bosonic mass $ m =0$, and the usual FL expression with $\bar{\Sigma}_1 \propto i \, \nu_n$ for $m \neq0$ (see, for example, Ref.~\cite{Wolfle_RMP}). 

As mentioned above, these results are analogous to those for an Ising-nematic QCP in a metal. The difference here is that the QCP is approached from the ordered state, rather than from the disordered state. More importantly, in our case, it is not the gap in the amplitude fluctuations, but the small mass of the pseudo-Goldstone mode associated with phase fluctuations that restores the FL behaviour, as one moves away from the QCP. These phase fluctuations, in turn, couple to the fermionic degrees of freedom via a Yukawa-like coupling, rather than a gradient-like coupling (typical for phonons). The key point is that because the pseudo-Goldstone behaviour arises from a dangerously irrelevant variable, its relevant critical exponent $\xi'$ is different from the critical exponent $\xi$ associated with the amplitude fluctuations. 

\section{Discussion and conclusions}

\label{secend}

Our calculations with the patch model, assuming convex Fermi surfaces with antipodal patches with parallel tangent vectors, show that $\Omega^{*}\sim\lambda^{3/2}$. In other words, the energy scale $\Omega^{*}$, associated with the NFL-to-FL crossover, is directly related to the dangerously irrelevant coupling constant $\lambda$ of the six-state clock model. This has important consequences for the energy range in which the NFL is expected to be observed in realistic settings. In the classical 3D $Z_{6}$ clock model, it is known that the dangerously irrelevant variable $\lambda$ introduces a new length scale $\xi'$ in the ordered phase \cite{Sandvik2007,Okubo2015,Leonard2015,Podolski2016}. It is only beyond this length scale that the discrete nature of the broken symmetry is manifested; below it, the system essentially behaves as if it were in the ordered state of the XY model. Like the standard correlation length $\xi$, which is associated with fluctuations of
the amplitude mode, $\xi'$ also diverges upon approaching the QCP from the ordered state. However, its critical exponent $\nu'$ is larger than the XY critical exponent $\nu$, implying that $\xi'\gg\xi$ as the QCP is approached. As a result, there is a wide range of length scales for which the ordered state is similar to that of the XY model. 

Applying these results to our quantum model, we therefore expect a wide energy range for which the fermionic self-energy displays the same behaviour as fermions coupled to a hypothetical XY nematic order parameter, i.e., the NFL behaviour $\Sigma\sim i \,\mathrm{sgn}(\nu_n) \left|\nu_{n}\right|^{2/3}$. Thus, the actual crossover energy scale $\Omega^{*}$ should be very small compared with other energy scales of the problem. This analysis suggests that the valley-polarized nematic state in a triangular lattice is a promising candidate to display the strange metallic behaviour predicted originally for the ``ideal'' (i.e., hypothetically uncoupled from the lattice) XY nematic phase in the square lattice \cite{Oganesyan01}. 

It is important to point out a caveat with this analysis. Although the aforementioned critical behaviour of the $Z_{6}$ clock model has been verified by Monte Carlo simulations, for both the 3D classical case and the 2D quantum case \cite{Sandvik2021}, the impact of the coupling to the fermions remains to be determined. The results of our patch model calculations for the bosonic self-energy show the emergence of Landau damping in the dynamics of the phase fluctuations, which is expected
to change the universality class of the QCP --- and the value of the exponent $\nu$ --- from 3D XY to Gaussian, due to the reduction of the upper critical dimension. The impact of Landau damping on the crossover exponent $\nu'$ is a topic that deserves further investigation, particularly since even in the purely bosonic case, there are different proposals for the scaling expression for $\nu'$ (see Ref.~\cite{Sandvik2021} and references therein).

We also emphasize the fact that our results have been derived for $T=0$. Experimentally, however, NFL behaviour is often probed at nonzero temperatures. It is therefore important to determine whether the NFL behaviour of the self-energy persists at a small nonzero temperature. At first sight, this may seem difficult, since in the classical 2D $Z_{6}$ clock model, the $\lambda$-term
is a relevant perturbation. In fact, as discussed in Sec.~\ref{secmodel}, the system in 2D displays two Kosterlitz-Thouless transitions, with crossover temperature scales of $T_{\mathrm{KT},1}$ and $T_{\mathrm{KT},2}$, with $Z_{6}$ symmetry-breaking setting in below $T_{\mathrm{KT},2}$ \cite{Jose1977}. However, a more in-depth analysis, as outlined in Ref.~\cite{Podolski2016}, indicates that as the QCP is approached, a new
crossover temperature $T^{*}<T_{\mathrm{KT},2}$ emerges, below which the ordered state is governed by the QCP (rather than the thermal transition). Not surprisingly, the emergence of $T^{*}$ is rooted on the existence of the dangerously irrelevant perturbation along the $T=0$ axis.
Therefore, as long as $\Omega^{*}<T^{*}$, the NFL behaviour is expected to be manifested at nonzero temperatures. Whether and how it is manifested in resistivity measurements, which are the primary tools to probe NFL behaviour, require further investigations beyond the scope of this paper. One of the issues involved is that the quasiparticle inverse lifetime, which can be obtained directly from the self-energy, is different from the actual transport scattering rate, which is hardly affected by small-angle scattering processes \cite{Maslov2011,Hartnoll2014,Carvalho2019}. 

An obvious candidate to display a valley-polarized nematic state is twisted bilayer graphene and, more broadly, twisted moir\'e systems.  Experimentally, as we showed in this paper, a valley-polarized nematic state would be manifested primarily as in-plane orbital ferromagnetism, breaking threefold rotation, twofold rotation, and time-reversal symmetries. While several experiments have reported evidence for out-of-plane orbital ferromagnetism \cite{Sharpe19,Efetov19,Young19,Tschirhart2021}, it remains to be seen whether there are regions in the phase diagram where the magnetic moments point in-plane \cite{Berg2022}. An important property of the valley-polarized nematic state is that the Dirac points remain protected, albeit displaced from the $K$ point, since the combined $C_{2z} \mathcal{T}$ operation remains a symmetry of the system.

A somewhat related type of order, which has also been proposed to be realized in twisted bilayer graphene and other systems with higher-order Van Hove singularities \cite{Chichinadze2020,Classen2020}, is the \emph{spin-polarized nematic order} \cite{Kivelson_RMP,Wu_Fradkin2007,Fischer_Kim2011}. It is described by an order parameter of the form $\overrightarrow{\varphi}=\left(\overrightarrow{\varphi}_{1},\,\overrightarrow{\varphi}_{2}\right)$,
where the indices denote the two $d$-wave components associated with the irreducible representation $E_{2}$ of the point group $\mathrm{D}_{6}$. The arrows denote that these quantities transform as vectors in spin space. The main difference between
$\overrightarrow{\varphi}$ and the valley-polarized
nematic state is that the spin-polarized nematic state does not break the $C_{2z}$ symmetry. It is therefore interesting to ask whether our results would also apply for this phase. The main issue is that $\overrightarrow{\varphi}$ is not described by a six-state clock model, since an additional quartic term is present in the action (see Ref.~\cite{Classen2020}), which goes as:
\begin{align}
S_{\vec{\varphi}}\sim\left(\vec{\varphi}_{1}\cdot\vec{\varphi}_{2}\right)^{2}-\left|\vec{\varphi}_{1}\right|^{2}\left|\vec{\varphi}_{2}\right|^{2} \,.
\label{spin_polarized}
\end{align}
However, if spin-orbit coupling is present in such a way that $\overrightarrow{\varphi}$ becomes polarized along the $z$-axis, this additional term vanishes. The resulting order parameter $\varphi^{z}=\left(\varphi_{1}^{z},\,\varphi_{2}^{z}\right)$ transforms as the $E_{2}^{-}$ irreducible representation, and its corresponding action is the same as Eq.~(\ref{actionboson0}), i.e., a six-state clock model. Moreover, the coupling to the fermions has
the same form as in Eq.~(\ref{S_bf}), with $a$ now denoting the spin projection, rather than the valley quantum number. Consequently, we also expect an NFL-to-FL crossover inside the Ising spin-polarized nematic state. 

In summary, we presented a phenomenological model for the emergence of valley-polarized nematic order in twisted moir\'e systems, which is manifested as in-plane orbital ferromagnetism. More broadly, we showed that when a metallic system undergoes a quantum phase transition to a valley-polarized nematic state, the electronic self-energy at $T=0$ in the ordered state displays a crossover from the FL behaviour (at very low energies) to NFL behaviour (at low-to-moderate
energies). This phenomenon is a consequence of the six-state-clock ($Z_{6}$) symmetry of the valley-polarized nematic order parameter, which implies the existence of a pseudo-Goldstone mode in the ordered state, and of a Yukawa-like coupling between the phase mode and the
itinerant electron density. The existence of the pseudo-Goldstone mode arises, despite the discrete nature of the broken symmetry, because the anisotropic $\lambda$-term in the bosonic action [cf. Eq.~(\ref{actionboson0})], which lowers the continuous O(2) symmetry to $Z_{6}$, is a dangerously irrelevant perturbation. Our results thus provide an interesting route to realize NFL behaviour in twisted moir\'e systems.

\begin{acknowledgments}
We thank A. Chakraborty, S.-S. Lee, A. Sandvik, and C. Xu for fruitful discussions.
RMF was supported by the U. S. Department of Energy, Office
of Science, Basic Energy Sciences, Materials Sciences and Engineering
Division, under Award No. DE-SC0020045.
\end{acknowledgments}


\bibliography{ref}

\end{document}